\documentclass[twocolumn,showpacs,amsmath,amssymb,aps,prb,superscriptaddress]{revtex4}
\usepackage{graphicx}
\usepackage{dcolumn}
\usepackage{bm}
\usepackage{multirow}
\newcommand{\comment}[1]{}
\newcommand{\qsgw}{QS\emph{GW}}
\newcommand{\bfq}{{\bf q}}

\begin{document}
\title{First-principles study of phase stability of Gd-doped EuO and EuS}
\author{J. M. An}
\email{anj_777@msn.com}
\affiliation{Department of Physics and Astronomy, Nebraska Center for
Materials and Nanoscience, University of Nebraska, Lincoln, NE 68588, USA}
\author{S. V. Barabash}
\author{V. Ozolins}
\affiliation{Department of Materials Science and
Engineering, University of California, Los Angeles, CA 90095-1595, USA}
\author{M. van Schilfgaarde}
\affiliation{Arizona State University, Tempe, Arizona 85284, USA}
\author{K. D. Belashchenko}
\affiliation{Department of Physics and Astronomy, Nebraska Center for
Materials and Nanoscience, University of Nebraska, Lincoln, NE 68588, USA}

\date{\today}

\begin{abstract}
Phase diagrams of isoelectronic  Eu$_{1-x}$Gd$_x$O and Eu$_{1-x}$Gd$_{x}$S
quasi-binary alloy systems are constructed using first-principles
calculations combined with the standard cluster expansion approach and
Monte-Carlo simulations. The oxide system has a wide miscibility
gap on the Gd-rich side but forms ordered compounds on the Eu-rich side,
exhibiting a deep asymmetric convex hull in the formation enthalpy diagram.
The sulfide system has no stable compounds. The large difference in the
formation enthalpies of the oxide and sulfide compounds is due to the
contribution of local lattice relaxation, which is sensitive to the anion
size. The solubility of Gd in both EuO and EuS is in the range of 10-20\% at
room temperature and quickly increases at higher temperatures, indicating
that highly doped disordered solid solutions can be produced without the
precipitation of secondary phases. We also predict that rocksalt GdO can be
stabilized under appropriate experimental conditions.

\end{abstract}
 \maketitle

\section{Introduction}

The unique properties of gadolinium-doped europium chalcogenides make them
attractive for spintronic and neutron detection applications. Doped EuO
undergoes a spectacular metal-insulator transition near its Curie temperature
$T_c$, which is accompanied by huge magnetoresistance. In some samples, the
resistivity changes by up to 13 orders of magnitude upon changing the
temperature, \cite{MIT,MIT3,MIT2} or by up to 6 orders of magnitude upon the
application of an external magnetic field. \cite{eugdo3,MIT3} Doped EuO can
be epitaxially grown on Si and GaN substrates, and it demonstrates a very
high spin polarization of the conducting electrons in the ferromagnetic (FM)
state. \cite{eugdo3} This half-metallic behavior suggests applications of
doped EuO as a spin injector material in spintronic heterostructures. On the
other hand, Gd-doped semiconductors are appealing as neutron-absorbing
materials for solid-state neutron detection technology due to a very high
neutron absorption cross-section of the $^{157}$Gd isotope.
\cite{gd2,gd1,Ketsman,Schultz}

The potential applications of Gd-doped Eu chalcogenides depend on their phase
stability. No phase diagrams are available for Eu-Gd oxides; the data for
Eu-Gd sulfides is limited to the EuS-Gd$_2$S$_3$ isoplethal
section.\cite{eliseev89} The experimental studies of pure and Gd-doped EuO
and EuS give only indirect information about the phase stability while
focusing on other properties, which we briefly review here. Pure EuO is a
rare FM insulator with a rocksalt structure, whose optical absorption gap
increases monotonically from 0.95 eV at 0 K to 1.12 eV at 300 K \cite{euo1}.
A divalent Eu ion in EuO has the $^{8}S_{7/2}$ ground state configuration and
the magnetic moment of 7$\mu_{\mathrm{B}}$ due to the half-filled 4$f$ shell.
The Curie temperature rises sharply with Gd doping from $T_c=69$ K in pure
EuO up to $T_c=170$ K at optimal doping of about 4\%.
\cite{mag1,mag2,eugdo1,eugdo2} The miscibility of Gd in EuO is unknown but is
expected to be finite, particularly because GdO has not been observed in the
rocksalt structure. In fact, the common Gd oxide has Gd$_2$O$_3$
stoichiometry (crystallizing in three different phases \cite{sesquioxide}),
but a tentative observation of zincblende GdO has also been reported.
\cite{Bist} On the other hand, both EuS and GdS are stable in the rocksalt
structure, and they can form a continuous range of rocksalt solid solutions
at all concentrations.\cite{eugds1} EuS is an insulator with the absorption
gap of 1.64 eV, which is FM below $T_c=19$ K;\cite{euo1} GdS is an AFM metal
with a N\'eel temperature of 58 K. \cite{eugds1}

Several recent first-principles studies
\cite{euo0,Larson,GhoshEuO,Ingle08,EuOferroel} have focused on the electronic
structure, magnetic interaction, and other properties of pure Eu
monochalcogenides. The interplay between the magnetic ordering, spectral, and
transport properties of doped EuO was also studied using model Hamiltonians.
\cite{Nolting1,Nolting2,Arnold} The structural phase stability of the
(Eu,Gd)O and (Eu,Gd)S quasi-binary alloys thus remains unexplored.

The purpose of the present study is to give a theoretical perspective of the
phase identity, stability, miscibility and other properties of Gd-doped EuO
and EuS along the EuO-GdO and EuS-GdS isoplethal sections. Based on the
comparison with the self-consistent quasiparticle GW calculations and with
available experimental data, we adopt the generalized gradient approximation
(GGA) with the Hubbard $U$ corrections applied only to the rare-earth $4f$
orbitals. We then apply the first-principles cluster expansion technique and
subsequent Monte Carlo simulations to construct the phase diagrams of
(Eu,Gd)O and (Eu,Gd)S quasi-binary alloys. We find that, despite the
isovalency of the two alloy systems, the resulting phase diagrams are quite
different. In particular, we predict that the oxide system has two
yet-unobserved ordered phases with 1:1 and 1:2 Gd-to-Eu ratios, which
become thermodynamically stable below $\sim900$ and $\sim500$ K;
moreover, we predict that rocksalt GdO can be stabilized
in a narrow range of oxygen pressures. On the other hand,
such 1:1 and 1:2 phases do not appear in the (Eu,Gd)S system,
and moreover a different ordered phase with a 2:1 Gd:Eu ratio
is very near the tie-line of the end compounds EuS and GdS.
We further analyze the role of the chemical and deformation-mediated
interactions and find that the qualitative difference between the oxide and
sulfide systems is mainly due to the contribution of local anion relaxations.

The paper is organized as follows. The methodological and computational
details are described in Section \ref{sec:approach}. Section
\ref{sec:BandStructure} presents the calculations of band structure and
elastic properties of end compounds. Comparison with experiment and with $GW$
calculations serves to justify the adopted GGA$+U$ approach. Section
\ref{sec:Magnetism} discusses the magnetic ordering energies and their
relevance to phase stability. The configurational Hamiltonians are described
in Section \ref{sec:CE}, followed by the calculations of phase diagrams in
Section \ref{sec:stability}. The anion-mediated deformational interaction
mechanism is discussed in Section \ref{sec:mechanism}, and finally we
summarize in Section \ref{sec:summary}. Some technical details, including the
extraction of paramagnetic formation enthalpies and the structural
information for the predicted compounds, are included in the Appendices.

\section{Computational approach}\label{sec:approach}

\subsection{Total energy calculations}

Total energy calculations for all ordered compounds in this study were
performed using the projected augmented wave (PAW) method
\cite{Bloechl,VASP-PAW} and generalized gradient approximation (GGA) of
Perdew-Burke-Ernzerhof, \cite{PBE} with the Hubbard $U$ correction
\cite{ldau} for the $4f$ orbitals, as implemented in the VASP package.
\cite{VASP} An energy cutoff of 500 eV was used for the plane-wave expansion
of wave functions, and total energies were converged to within a few meV per
atom with the density of the k-point mesh no lower than 0.01 \AA$^{-3}$
including the $\Gamma$ point for Brillouin-zone sampling.

From the energetics of the individual atomic levels, it is clear that in both
oxides and sulfides, the $4f$ bands are half-filled, whether originating from
Eu or Gd, whereas the binding energy of the Gd $4f$ states is much larger
compared to divalent Eu. Electron doping through the addition of Gd fills the
conduction-band states, which are spin-split by the exchange interaction with
the $4f$ shell. The shallow core $5s^2$ and $5p^6$ states on both Eu and Gd
are included in the valence basis set. The half-filled and strongly-localized
$4f$ orbitals in both Eu and Gd were treated within the GGA$+U$ approach.
\cite{ldau} To calculate the value of $J=0.6$ eV, we used the constrained
occupation method \cite{constr1,constr2} by considering the $4f$ states as an
open-core shell and finding the total energy difference between the
4$f^{7}_{\uparrow}$ 4$f^{0}_{\downarrow}$ and 4$f^{6}_{\uparrow}$
4$f^{1}_{\downarrow}$ states. This calculation was performed using the
full-potential linear augmented plane wave (FLAPW) method implemented in the
FLEUR package. \cite{FLEUR} The value $U=5.3$ eV calculated in a similar way,
when used in the GGA$+U$ calculation, leads to the $4f$ states being too
shallow with respect to the conduction band of EuO. This discrepancy is due
to the underestimation by GGA of the intrinsic insulating gap between the
O-$2p$ and Eu-$5d$ states. This underestimation (which is not sensitive to
$U$) brings the O-$2p$ states too close to the Eu $4f$ states; their
hybridization-induced repulsion pushes the $4f$ states up and reduces the
band gap. Therefore, we adjusted the value of $U$ empirically and found that
$U = 7.5$ eV results in good agreement with optical absorption
\cite{euo3,euo4} and photoemission \cite{Eastman} measurements, as
illustrated in section \ref{sec:BandStructure}. The Hubbard $U$ corrections
are {\em not} used for the $5d$ orbitals on Eu and Gd, as justified below in
sections \ref{sec:GW} and \ref{sec:BandStructure}.

The proper treatment of the chemical disorder in many $f$ and $d$ electron
systems is notoriously difficult, because of the dependence of the structural
energy on the $f$ (or $d$) orbital orientation, and due to the large spurious
self-interacion present in the popular GGA+U versions for $f$ (or
non-time-reversible $d$) orbitals. \cite{FeiSIC} Fortunately, this is not an
issue in (Eu,Gd)O and (Eu,Gd)S, since the half-filled $f$ shells do not
exhibit such an orientational energy dependence. Indeed, for all considered
cases, we found the lowest energy electronic configurations to correspond to
exactly seven co-aligned $f$ electrons on each cation atom, forming a
rotationally-invariant shell. At the same time, it was not unusual for our
early test calculations to lead to electronic states with different
$f$-electron occupation, with energies higher by 0.15\dots 3.0 eV/cation.
Some of such higher-energy states were clearly identifiable as having a Gd
electronic configuration of $4f^8 5d^0$ instead of $4f^7 5d^1$. We found that
the appearance of such states was either a failure of the residual
minimization method (RMM), \cite{RMM} or an artifact of an insufficient
initial $f$-electron spin polarization. By using the Davidson minimization
algorithm\cite{Davidson} and by assigning the initial on-site spin
polarization of 10 $\mu_B$ per cation (which is partially assigned to the $d$
and $s$ electrons, resulting in the desired occupation of 7 co-aligned $f$
electrons), we could avoid such high-energy electronic configurations. The
only remaining degree of freedom for the final electronic state corresponds
to the magnetic ordering of the fully polarized cations, which is well
controlled by the signs of the initial on-site spin polarizations and is
further discussed in Section \ref {sec:Magnetism}.

The spin-orbit corrections are often appreciable for the band energies of
heavy-element compounds at high-symmetry k-points, but are typically minor for
the formation energies, which include contributions averaged over different
bands in the entire Brillouin zone. Indeed, we  checked the effects of the
spin-orbit coupling on the formation enthalpies $\Delta H_\mathrm{GGA}$ for
the key ordered structures that we identify below in Sec.\ref{sec:stability}
as the stable ground state compounds. We found that the $\Delta
H_\mathrm{GGA}$ change by at most 2 meV/cation (see Table \ref{enthalpies}
below). We therefore performed all our routine calculations in the
scalar-relativistic approximation, without including the spin-orbit coupling.

\subsection{Benchmark $GW$ calculations}
\label{sec:GW}

It has been suggested by some authors \cite{GhoshEuO,Larson} that the
addition of Hubbard corrections for the empty $5d$ states may be necessary
for the correct description of the conduction band, and in particular for the
determination of the character of the band gap (direct or indirect) of
europium chalcogenides. However, these corrections are arbitrary unless a
reliable benchmark is used to select the Hubbard parameters. We resolve this
issue by calculating the band structure of EuO using the quasiparticle
self-consistent $GW$ (\qsgw) approximation. Here, we discuss the \qsgw
methodology, and later in Section \ref{sec:BandStructure} we use \qsgw\ to
show that the conduction band structure comes out almost exactly right in the
GGA$+U$ calculation with $U$ applied \emph{only} to the $4f$ orbitals.

\qsgw\ has been shown to be a reliable predictor of materials properties for
a wide range of compounds composed of elements throughout the periodic
table.~\cite{Kotani07,Faleev04,mark06qsgw,Chantis06a,Chantis07a}
Nevertheless, prior experience has revealed certain kinds of systematic
errors. The correction of these errors makes minor adjustments to weakly
correlated materials systems, and somewhat stronger adjustments for more
correlated materials.  There are two highly systematic errors that affect the
band structure of EuO.

First, bandgaps in semiconductors such as GaAs, and insulators
such as SrTiO$_{3}$ are systematically overestimated a little. The same
effect is seen in the $spd$ subsystem of EuO.  Second, shallow
core-like levels, such as the highest occupied $d$ levels in
Zn,Cd,Cu,Ag,Au, and so on, are systematically predicted to be too
close to the Fermi level, typically by $\lesssim0.5$~eV.  This
error is seen in the $f$ subsystem of EuO, as we will discuss.

Both types of errors are highly systematic in $sp$ and $d$ systems, and
discrepancies with experiment in $4f$ compounds are consistent with these
errors.\cite{Chantis07a} To a large extent, the first error can be simply
explained through the random phase approximation (RPA) to the screened
interaction $W$, which can be understood as follows.  The RPA bubble diagrams
do not include electron-hole interactions in intermediate states in the
calculation of the irreducible polarizability $\Pi(\bfq,\omega)$ and thus the
dielectric function $\epsilon(\bfq,\omega)$. Short-range attractive
(electron-hole) interactions induce the red shifts in
$\mathop{\mathrm{Im}}\epsilon(\bfq,\omega)$ at energies well above the
fundamental bandgap; see e.\ g.\ Fig.\ 6 in Ref.\ \onlinecite{Kotani07}.
Ladder diagrams are sufficient to remedy most of the important errors in
$\Pi(\bfq,\omega)$, as was demonstrated rather convincingly in
Cu$_2$O.\cite{Bruneval06b}

Inclusion of these contributions increases the static dielectric constant
$\epsilon_{\infty}$, as can be readily seen through the Kramers-Kronig
formula relating the real and imaginary parts of $\epsilon$.  Remarkably,
$\epsilon_{\infty}$ calculated by the RPA in \qsgw\ is underestimated by a
nearly \emph{universal factor} of 0.8, for many kinds of insulators and
semiconductors,\cite{mark06qsgw} including transition metal oxides such as
Cu$_{2}$O, \cite{Kotani07}, SrTiO$_{3}$ \cite{Chambers04}, CeO$_{2}$, and
\emph{sp} semiconductors.\cite{Shishkin07} Because $\epsilon$ is
systematically \emph{underestimated}, $W=\epsilon^{-1} v$ (where $v$ is the
Coulomb interaction), the self-energy $\Sigma=-iGW$, and the quasiparticle
excitation energies are systematically \emph{overestimated}.

The second kind of error cannot be explained in this way.  \qsgw\ pushes down
the semicore $d$ level in Zn (or $f$ level in Eu) rather strongly relative to
the LDA; however, the shift is slightly too small to agree with experiment.
As we have noted, $W$ calculated by \qsgw\ is already too large: reducing $W$
reduces this correction.  This implies that the error should be attributed to
the other approximation in $GW$ theory, namely the omission of the vertex
$\Gamma$ in the formally exact self-energy, $GW\Gamma$.

Both kinds of corrections to the $GW$ approximation are difficult to carry
out in practice.  As for the correction to $\Pi$, we have noted that simply
scaling $\Sigma$ by 0.8 largely undoes this error, in a wide range of
systems.  We make such a scaling here, to correct the $spd$ subsystem.

Whether or not $\Sigma$ is scaled, the Eu $d$-$f$ gap is too small: when
$\Sigma$ is scaled it comes out approximately zero, in contrast to the
observed gap of about 1 eV.  It is expected that the vertex in $GW\Gamma$
will largely just induce a shift in the Eu $f$ state.  Anticipating this, we
included an \emph{ad hoc} addition to the \qsgw\ potential for EuO,
essentially doing a \qsgw+$U$ calculation with $U=0.816$~eV.  The value of
$U$ is adjusted to make the $d$-$f$ gap coincide with the experimental
number. With these corrections, we anticipate the \qsgw\ method to yield
highly accurate band structures, discussed below in Sec.
\ref{sec:BandStructure}.

\subsection{Cluster expansion and Monte Carlo simulations}
\label{sec:CEmethod}

To identify the thermodynamically stable phases and their range of stability,
the standard cluster expansion (CE) formalism\cite{sanchezCE,Zunger} coupled
with the ground state search and Monte Carlo simulations were employed, using
the routines implemented in the Alloy-Theoretic Automated Toolkit (ATAT)
package. \cite{atat1,atat2} We consider quasi-binary substitutional alloys,
assuming that the anion sublattice is fully occupied by the chalcogenide
atoms of one kind (either O or S). Throughout the paper, we use the terms
``structure'' and ``ordering'' to refer specifically to the ordering of Eu
and Gd cations within their own (fcc) sublattice, unless indicated otherwise.
Our CEs cover FM compounds, with other cases considered separately, as
detailed in Sec. \ref{sec:Magnetism}. The cell size and shape, as well as all
the atomic positions, were fully relaxed for all structures using the
conjugate gradient algorithm, starting from the ideal rocksalt lattice. In
view of prohibitive computational cost, we did not consider the phonon
contribution to the formation enthalpies, which can somewhat modify the phase
diagrams at elevated temperatures.

The relaxed formation enthalpies are parameterized by a configurational
cluster-expansion Hamiltonian $\Delta H_\mathrm{CE}$:
\begin{equation}
\Delta
H_\mathrm{CE}(\boldsymbol{\sigma})=\sum_{f}J_{f}D_{f}\bar{\Pi}(\boldsymbol{\sigma})~,
\label{eq1}
\end{equation}
where the occupational degrees of freedom are described by a configurational
vector $\boldsymbol{\sigma}$ (a particular decoration of the cation sites of the
rocksalt lattice by Gd and Eu atoms), $J_{f}$ is called the effective cluster
interaction (ECI) for a cluster figure $f$ with $D_{f}$ as the figure's
symmetry degeneracy per site, and $\bar{\Pi}(\boldsymbol{\sigma})$ is the
configuration-dependent correlation function in the interaction cluster.
In practice, a finite number
of terms $n_f$ is kept in the expansion (\ref{eq1}), and the
expansion becomes exact\cite{sanchezCE} as $n_f\to\infty$.

The ECI values $J_{f}$ are determined by fitting to a set of $N_\mathrm{in}$
``input'' formation enthalpies $\Delta
H_\mathrm{GGA}(\boldsymbol{\sigma}_\mathrm{in})$. $N_\mathrm{in}$ was
iteratively  increased by performing GGA$+U$ calculations for new structures
$\boldsymbol{\sigma}_\mathrm{in}$ based on the CE predictions, until a
desired CE accuracy was reached, in particular establishing an agreement
between the ground state predictions of the final CE and GGA$+U$. (For the
oxide system such full consistency was established only within a target
concentration range discussed in Sec.\ref{sec:EuGdOstability}.) In order to
evaluate the predictive power of the cluster expansion, the ``leave-one-out''
cross-validation (CV) score was calculated using the procedure implemented in
ATAT:\cite{atat1} for each $\bf\sigma_0$ out of the $N_\mathrm{in}$ input
structures, a separate fitting of the $J_{f}$ values was performed with that
one structure excluded from the fitted set. The actual energy $\Delta
H_\mathrm{GGA}(\boldsymbol{\sigma_0})$ of the excluded structure was then
compared with the prediction $\Delta H_\mathrm{CE}^{\mathrm{no}\,\sigma_0}
(\boldsymbol{\sigma_0})$ of this ``leave-one-out'' fitting, and the
difference was averaged over all the $N_\mathrm{in}$ choices of $\bf\sigma_0$
. Unlike the conventional mean-squares fit error, which monotonously
decreases upon increasing $n_f$, the CV score is designed to measure the {\em
predictive} power of the cluster expansion and has a minimum for a finite
$n_f$ value, diverging if $n_f$ becomes too large.

Once the ground states for a range of concentrations have been identified,
the ones that are stable at $T=0$ K are determined by the convexity condition;
the given structure at concentration $x$ is stable if it lies below any
straight line connecting other compounds at concentrations $x_1<x$ and
$x_2>x$. (The convex hull of the set of points in the $\Delta H$ \emph{vs}
$x$ plot represents the full range of enthalpies that may be achieved by the
system. Stable compounds are those that form the vertices on the lower
boundary of this convex hull.) For each
stable compound one can define its ``energetic depth'' $\delta$, i.\ e.\ the
amount by which its energy would increase if it were decomposed into two
stable compounds that are closest to it in concentration.

The phase diagrams were computed using semi-grand canonical Monte Carlo (MC)
simulations (i.~e. with the varying number of Eu and Gd cations) and the
Metropolis algorithm implemented in ATAT. For the final phase diagram
construction, we used an $18\times 18\times 18$ supercell based on primitive
fcc translations (5832 cations in the simulation box). In the case of
(Eu,Gd)O, we estimated the effects of the finite size and the
commensurability with other ground state structures by also using $12\times
12\times 12$ and $15\times 15\times 15$  simulation boxes. We found that
the $18\times 18\times 18$ box was commensurate with all the ground-state
structures that have ordering temperature $T_{ord}$ above 400 K, although
this did not hold for some structures stable at lower $T$ (either identified
by the CE ground state search or directly observed in MC simulations). All
such low-$T_{ord}$ structures in the oxides are limited to the Eu-rich composition
range indicated below in the phase diagram. A series of simulations was performed
at fixed chemical potentials with temperatures varying in 2--5 K increments.
The equilibration and sampling passes were done with 1000--5000 Monte Carlo
steps (flip attempts per site); longer runs of 10000 steps were performed for
particularly difficult regions. The phase boundaries were then found by
identifying the discontinuities (or cusps) in the dependence of the average
concentration and enthalpy on temperature.

\section{Binary Chalcogenides: properties and electronic structure}
\label{sec:BandStructure}

\begin{figure*}[hbt]
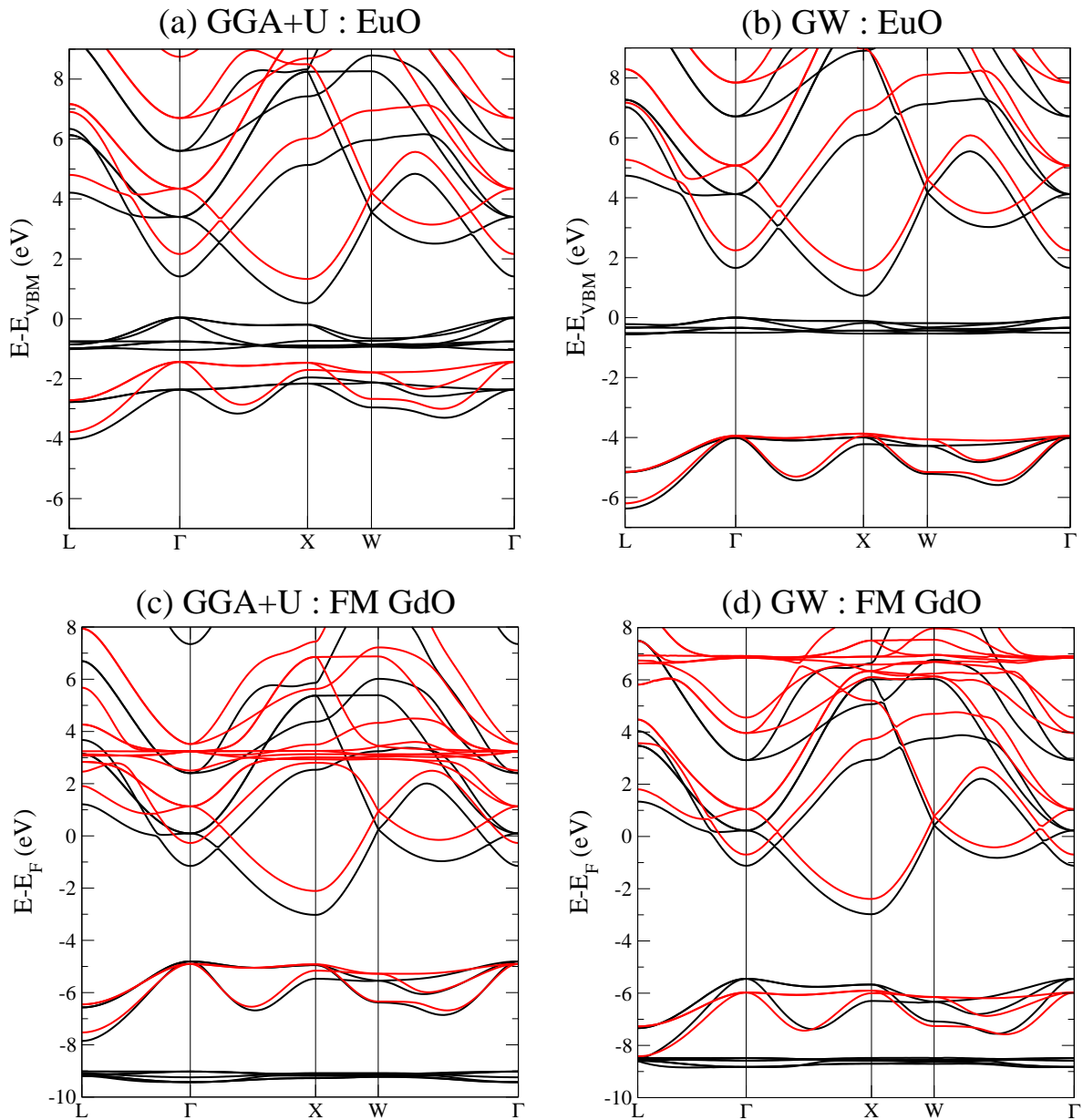

\centering
\begin{minipage}{17cm}
\begin{minipage}{8cm}
\includegraphics[width=7.5cm,angle=0,clip]{fig1a.eps}
\vspace{0.5cm}
\end{minipage}
\begin{minipage}{8cm}
\includegraphics[width=7.5cm,angle=0,clip]{fig1b.eps}
\vspace{0.5cm}
\end{minipage}
\begin{minipage}{8cm}
\includegraphics[width=7.5cm,angle=0,clip]{fig1c.eps}
\end{minipage}
\begin{minipage}{8cm}
\includegraphics[width=7.5cm,angle=0,clip]{fig1d.eps}
\end{minipage}
\end{minipage}
\caption{(Color online) Electronic band structure of EuO calculated (a) in
GGA+U and (b) in \qsgw, and of ferromagnetic phase GdO, calculated by (c) GGA+U and
(d) \qsgw.
Black (red) curves correspond to the majority (minority) spin bands.
Energy is referenced either from the valence band maximum (VBM) or from the Fermi level (E$_F$).
In GdO, states near $E_{F}$ at $\Gamma$ are of Gd $d$ character; the band whose value
is $-$1.5~eV at $\Gamma$ is of $s$ character.  If correlations were strong,
the $d$ band would become narrow and possibly shift relative to the $s$ band.
As can be seen, GGA and \qsgw\ are very similar for these bands: GGA and
\qsgw\ differ mainly in the positions of the O $2p$ bands, at around
$-$6~eV.  That O $2p$ states shift downward relative to GGA (or LDA) seems to be a
universal property of oxide insulators.} \label{bands}
\end{figure*}

\begin{figure}[htb]
\centering
\includegraphics[width=8.5cm,angle=0,clip]{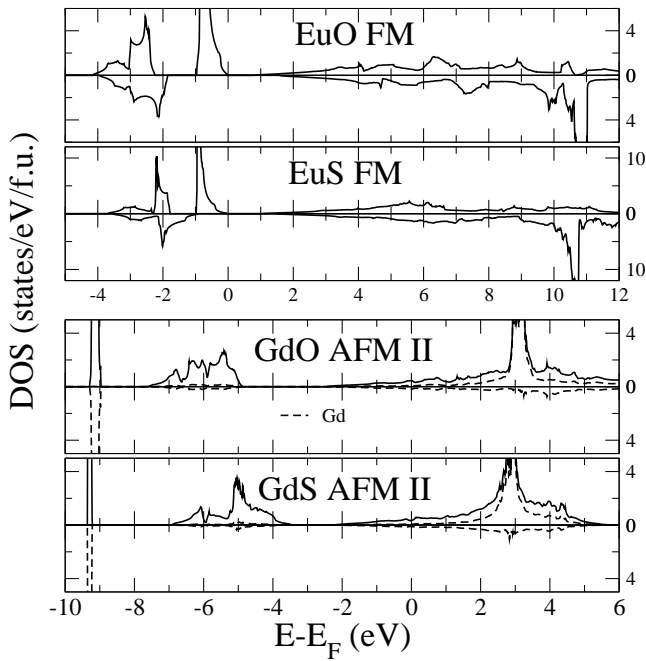}
\caption{Density of states (DOS) of FM EuO and EuS, AFM type II GdO, and AFM
type II GdS calculated in GGA+U. For the AFM II phases, the solid line shows
the total DOS including both the cation and the anion contributions, whereas
the dashed lines show the partial majority and minority-spin DOS from the Gd
cations. Energy is referenced from the Fermi level $E_F$.} \label{dos}
\end{figure}

Figure~\ref{bands} shows the electronic band structures of EuO and GdO
calculated using both GGA+U and \qsgw\ approaches. The densities of states of
all four end compounds calculated in GGA$+U$ are shown in Fig.\ \ref{dos}.

When Hubbard U is applied only to the rare-earth $4f$ states, the conduction
band structure of both EuO and GdO is in very good agreement between GGA$+U$
and \qsgw, the band gap being indirect. On the other hand, the addition of
the Hubbard $U$ correction to the Eu $5d$ orbitals proposed in Ref.\
\onlinecite{Larson} pushes the bands up at the $X$ point but not at $\Gamma$,
resulting in a direct band gap in disagreement with \qsgw. Therefore, as
mentioned above, we use GGA$+U$ with $U$ applied only to the rare-earth $4f$
states. Physically, this is reasonable because the $5d$ states of either Gd
or Eu are rather delocalized and have a sizable bandwidth, being thus
amenable to treatment within GGA.

The main difference between the GGA$+U$ and \qsgw\ band structures is the
position of the oxygen $p$ states in EuO. In GGA$+U$, they lie approximately
2.5 eV too high, resulting in a stronger hybridization of the majority-spin O
states with the occupied $4f$ states. This hybridization leads to a repulsion
of these states from each other, and to a large spin splitting of the oxygen
$p$ states. The \qsgw\ results are in very good agreement with photoemission
measurements.\cite{Eastman} By contrast, for GdO the oxygen $p$ states come
out only a little too shallow. This is reasonable, because the gap between
the oxygen $p$ and rare earth $5d$ states in GdO is not accompanied by a
discontinuity of occupation numbers; the latter is responsible for the band
gap problem in semiconductors. We also note that \qsgw\ significantly
overestimates the energy of the unoccupied $4f$ states, which appears to be
its universal feature \cite{Chantis07a}. This error is immaterial for our
purposes.

The choice of $U=7.5$ eV and $J=0.6$ eV simultaneously produces the splitting
between the occupied and unoccupied $4f$ states of $U + 6J = 11.1$ eV in good
agreement with photoemission and inverse photoemission measurements for the
Eu metal,\cite{Lang} the optical band gap at the X point of 0.94 eV in EuO
consistent with the value of 0.95 eV measured at zero temperature,
\cite{euo3,euo4} and the equilibrium lattice constant $a=5.182$~\AA~ in good
agreement with the experimental value of $a=5.144$~\AA. The application of
the same $U$ and $J$ values to EuS leads to the optical gap at the X point of
1.52 eV consistent with the zero-temperature value of 1.51 eV. \cite{euo1}
Due to the similar nature of the half-filled $4f$ orbitals, these
semiempirical $U$ and $J$ values were applied to both Eu and Gd $4f$ states
in the oxide and sulfide systems. Both GdO and GdS are metallic, as expected.


Table \ref{elastic} includes the lattice parameters and elastic constants
calculated in the FM state, along with the available experimental data for
EuO, EuS, and GdS. The lattice constants are slightly underestimated by
approximately 1\%. The bulk moduli agree within the uncertainty of the
experimental data. The elastic constants obtained by ultrasonic measurements
have only been reported by one group. \cite{Shapira} The calculated $C_{11}$
constant agrees well with this measurement for both EuO and EuS. The $C_{44}$
constant is overestimated by approximately 45\%. The $C_{12}$ constant is
also overestimated, but comparison is hindered by a very large experimental
uncertainty. Note, however, that the bulk moduli obtained from these measured
elastic constants are among the lowest ones reported in the literature; it is
possible that these measurements are affected by off-stoichiometry.
Therefore, it is unclear whether the disagreement in $C_{44}$ and $C_{12}$ is
due to the inaccuracy of the GGA+U method or to experimental artifacts.

\begin{table}[t]
\caption{Calculated (this work, marked "theory") and experimental (when
available, marked "exper.") lattice parameters, elastic constants, and the
bulk modulus $B$ of the binary rocksalt compounds. All elastic constants are
given in Mbar units.}
\begin{ruledtabular}
\begin{tabular}{llccccc}
\multicolumn{2}{l}{Compound}
& $a$, \AA & $C_{11}$ & $C_{12}$ & $C_{44}$ & $B$ \\
\hline
\multirow{2}{*}{EuO}
        & theory & 5.18 & 1.89 & 0.62 & 0.78 & 1.04 \\
        & exper.   & 5.14 & 1.9\footnotemark[1]  & 0.42(8)\footnotemark[1]  & 0.54\footnotemark[1] & 0.92(6);\footnotemark[1] 1.10\footnotemark[2] \\
\hline
GdO
        & theory & 4.92 & 3.54 & 0.63 & 0.72 & 1.60 \\
\hline
\multirow{2}{*}{EuS}
        & theory & 6.02 & 1.47 & 0.24 & 0.39 & 0.65 \\
        & exper.    & 5.97 & 1.3\footnotemark[1]  & 0.11(8)\footnotemark[1]     & 0.27\footnotemark[1]     & 0.51;\footnotemark[1] 0.61;\footnotemark[2] 0.72\footnotemark[3] \\
\hline
\multirow{2}{*}{GdS}
        & theory & 5.62 & 3.06 & 0.33 & 0.36 & 1.24 \\
        & exper.    & 5.56 &      &      &      & 1.20 \\
\end{tabular}
\end{ruledtabular}
\footnotetext[1]{Ref. \onlinecite{Shapira}} \footnotetext[2]{Ref.
\onlinecite{Jayaraman}} \footnotetext[3]{Ref. \onlinecite{Goncharenko}}
\label{elastic}
\end{table}

\section{Magnetic ordering}
\label{sec:Magnetism}

The magnetic ordering temperatures of pure EuO, EuS and GdS (69, 19, and 58
K, respectively), as well as for the entire range of solid solutions, are
well below room temperature. Therefore, all structural phase equilibria
involve paramagnetic (PM) phases with randomly oriented local moments on the
Eu and Gd atoms. In principle, the PM enthalpies for the input structures can
be estimated by fitting a number of magnetic configurations to a Heisenberg
Hamiltonian and taking the constant term as the PM energy. However, doing
this for more than a few simple ordered compounds is computationally
prohibitive. Fortunately, relatively low magnetic ordering temperatures
suggest that the use of ground state formation enthalpies to study
configurational thermodynamics should not lead to large errors. Nevertheless,
in this section we perform a few checks and discuss the possible
modifications introduced in the cluster expansion by the replacement of FM
formation enthalpies by the PM ones.

We considered several collinear magnetic configurations in seven simple
(Eu,Gd)O and four (Eu,Gd)S compounds;\cite{note:AFMtests} these results are
summarized in Table \ref {table:FMAFM}. Like in the rest of our study,  the
scalar-relativistic approximation was used here. Where a comparison can be
made, we found good agreement with other published data.\cite{euo0,Larson}
For the (Eu,Gd)O system the FM state always has the lowest energy, except in
pure GdO where the AFM phase with the ordering vector along [111] is slightly
lower (by 0.6 meV/cation) than the FM phase.
This indicates that restricting our CE study to the FM compounds
is sufficient to yield an accurate description of the thermodynamic
phase stability at $T=0$ K.

One can argue that FM enthalpies
can also be used to predict the phase stability of oxides at higher
temperatures.
As mentioned in the introduction, the $T_c$ of Gd-doped EuO rises sharply to
about 170 K at $x\approx0.04$ and then slowly decreases (see e.g. Ref.\
\onlinecite{mag2}). However, the magnetization curves for these higher-$T_c$
alloys have a distinctive double-dome shape. Specifically, the magnetization
drops to a fairly small value at temperatures close to $T_c$ of pure Eu
(``main dome''), and extends a relatively weak tail up to the actual elevated
$T_c$. While the mechanisms of this behavior are not completely understood,
it is fair to assume that the dominant part of the PM-FM enthalpy difference
is released in the temperature range of the ``main dome.'' Since the
characteristic temperature of this feature does not strongly depend on the
doping level, we expect that the PM-FM enthalpy difference is a featureless
function of the concentration, and that it likely does not exceed 10
meV.\cite{note-pmfm} This conclusion is generally consistent with the data in
Table \ref{table:FMAFM}. Such correction is not likely to lead to significant
changes in the phase diagram, and we therefore use FM enthalpies for oxides
in the following.

In the sulfide system the situation is different for two reasons. First, as
we discuss in Section \ref{sec:EuGdSstability}, the FM CE predicts an ordered
EuGd$_2$S$_3$ structure with a high ordering temperature, which is only
marginally stable with respect to the pure FM EuS and GdS. Second, the
magnetic order changes from FM in EuO to AFM in GdS, and the latter has a
relatively high FM-AFM energy difference of $13.5$ meV/cation (see Table
\ref{table:FMAFM}). Therefore, for EuS, EuGd$_2$S$_3$, and GdS we have
estimated the PM-FM enthalpy differences, which are $-5.3$, $+1.3$, and
$-0.6$ meV/cation, respectively (see Appendix \ref{appendix:paramagnetic}).
These differences are sufficient to make the EuGd$_2$S$_3$ structure
marginally unstable. The effect on the phase diagram is considered in Section
\ref{sec:EuGdSstability}.

\begin{table}[t]
\caption{ Formation energies (in meV/cation) of select ordered (Eu,Gd)O
rocksalt compounds with different magnetic orderings including FM, AFM with
two different orientations of the layers of co-aligned spins, and
ferrimagnetic [where applies, with different arrangement of (111) layers of
up and down spins relative to the majority (A) and minority (B) cation
species]. Magnetic orderings requiring large cells were not considered, as
indicated by dashes.}
\begin{ruledtabular}
\begin{tabular}{lccccc}
\multirow{2}{*}{Compound} &
            \multirow{2}{*}{FM} &
                    \multicolumn{2}{c}{AFM} &
                                    \multicolumn{2}{c}{ferrimagnetic (111)}  \\
        &   & (111) & (001) &
                                    A$\uparrow$\,A$\uparrow$\,B$\downarrow $ &
                                       A$\uparrow$\,A$\downarrow $\,B$\uparrow$ \\
\hline
EuO         & 0      &12.2 &  15.8 & \multicolumn{2}{c}{ 8.2} \\
GdO     & 0      &-0.6   &  15.1 & \multicolumn{2}{c}{ 1.6} \\
EuS             & 0      & 0.1 &  3.4 & \multicolumn{2}{c}{ --- } \\
GdS             & 0      & -13.5   &  3.8 & \multicolumn{2}{c}{ --- } \\
L1$_0$ EuGdO$_2$
        &  60.6 & ---  & 78.1  & \multicolumn{2}{c}{  ---} \\
L1$_1$ EuGdO$_2$
        & -59.3 & -47.9 & ---   &   \multicolumn{2}{c}{  ---} \\
L1$_1$ EuGdS$_2$
                &   8.9 &  7.5 & ---   &   \multicolumn{2}{c}{  ---} \\
C6 Eu$_2$GdO$_3$
        & -41.5 & ---   & ---   & -36.6     & -33.5 \\
C6 EuGd$_2$O$_3$
        & -32.9     & ---   & ---   &  -21.4 & -17.0  \\
C6 EuGd$_2$S$_3$
                & -0.4          & ---   & ---   & -3.6   & -4.6  \\
\hline
\hline
\multirow{2}{*}{} &
            \multirow{2}{*}{FM} &
                    \multicolumn{4}{c}{ferrimagnetic (001)}  \\
        &  & \multicolumn{2}{c}{
            A$\uparrow$A$\downarrow $A$\uparrow$B$\downarrow $
            } &
                          A$\uparrow$A$\downarrow $A$\downarrow $B$\uparrow $ &
                              A$\uparrow$A$\downarrow $A$\downarrow $B$\downarrow $ \\
\hline
Z1 Eu$_3$GdO$_4$
        & 5.8 & \multicolumn{2}{c}{17.1}
                & 29.9  & 22.4\\
\end{tabular}
\end{ruledtabular}
\label{table:FMAFM}
\end{table}

\section{cluster expansions}\label{sec:CE}

In this section, we characterize the CEs
obtained at the end of the iterative CE construction
procedure. These CEs are used
in Sec.\ref{sec:stability} to evaluate the phase stability of Eu$_{1-x}$Gd$_{x}$O
and Eu$_{1-x}$Gd$_{x}$S rocksalt alloys.

\subsection{Eu$_{1-x}$Gd$_{x}$O}
\label{sec:EuGdOCE}

The initial input set for the self-consistent CE construction
included $N_\mathrm{in}^0=26$ structures including all the atoms up to 4 cations per
cell, except two such structures at EuGd$_3$O$_4$ composition.
The final CE for Eu$_{1-x}$Gd$_{x}$O has an input set of $N_\mathrm{in}=148$
structures (identified throughout the CE iterations as potential ground
states or otherwise as structures important for the CE accuracy), and uses an
ECI set of 8 pairs, 12 triplets and 16 quadruplets. The predictive power of
this CE is estimated by the CV score as 5.8 meV/cation, whereas the
root-mean-square fit error for the input structures is only 2.8 meV/cation.
The ground state search was performed among all the structures up to 30 atoms
per cell ($\sim2^{15}$ configurations).

\begin{figure}[!htb]
\centering
\includegraphics[width=8.5cm,angle=0,clip]{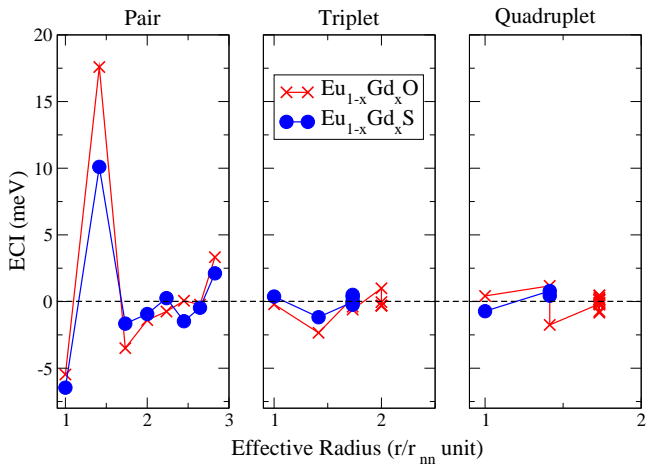}
\caption{(Color online) Three different types of ECI parameters as a function
of effective radius in units of the nearest-cation-neighbor distance $r_{nn}$
in the cluster expansions for Eu$_x$Gd$_{1-x}$O, denoted by red crosses, and
Eu$_x$Gd$_{1-x}$S, denoted by blue circles. } \label{ECIplot}
\end{figure}

The ECIs  values $J_f$ for Eu$_{1-x}$Gd$_{x}$O as a function of the effective
radius (the longest intersite distance in the cluster $f$) are shown in
Figure~\ref{ECIplot} as the crosses connected  by the red line. The leftmost
panel in Fig.~\ref{ECIplot} displays the pairwise ECIs up to the eighth
nearest-neighbor in the cation sublattice, and the right two panels
correspond to triplet and quadruplet ECIs, respectively. By far the largest
ECI is due to the second nearest-neighbor pair. It is positive, and its
magnitude (17.6 meV) is more than three times greater than that of the
negative first nearest-neighbor ECI ($-5.4$ meV). Overall, the pair
interactions are stronger than all other cluster ECIs. As explained in
Section \ref{sec:mechanism} below, the large positive ECI for second-nearest
neighbors is due to the significant energy gain from displacing O atoms
towards Gd in the Eu-O-Gd double-bond patterns along any of the [001]
directions. This ECI is the main driving force for the stabilization of the
L1$_1$ cation ordering that we find below in Sec.\ref {sec:EuGdOstability},
as indeed could be expected\cite{DucastelleBook} from a simple CE involving
only pair interactions $J_\mathrm{pair}^{nn}$ and $J_\mathrm{pair}^{2nn}$
between nearest-nighbor and second-nearest neighbor atoms of the same
magnitude as in our actual CE.

\subsection{Eu$_{1-x}$Gd$_{x}$S}
\label{sec:EuGdSCE}

The cluster expansion has been constructed with a CV score of 4.0 meV/cation
using $N_\mathrm{in}=49$ input structures and an ECI set of 8 pairs, 7
triplets, and 3 quadruplets. The set of structures used for ground state
search contained $\sim2^{10}$ configurations, and the root-mean-square fit
error for the input structures is only 2.2 meV/cation.

The pair and many-body ECIs for the oxide and sulfide systems are
qualitatively similar (see Fig.~\ref{ECIplot}). However,  we shall see in
Sec.\ref{sec:stability} that there is a striking difference in the
ground-state energetics, convex hull structures, and the phase diagrams of
the oxides and the sulfides. This difference stems from the quantitative
change in the ECIs: while the positive second nearest-neighbor pair ECI is
still the strongest one in the sulfide system, it is reduced by about 40\%
compared to the oxides. This reduction results in the destabilization of the
L1$_1$ cation ordering, as we discuss in Section \ref{sec:mechanism}.

\section{phase stability}
\label{sec:stability}

\subsection{Eu$_{1-x}$Gd$_{x}$O}
\label{sec:EuGdOstability}

\begin{figure}[!htb]
\centering
\includegraphics[width=0.45\textwidth,angle=0,clip]{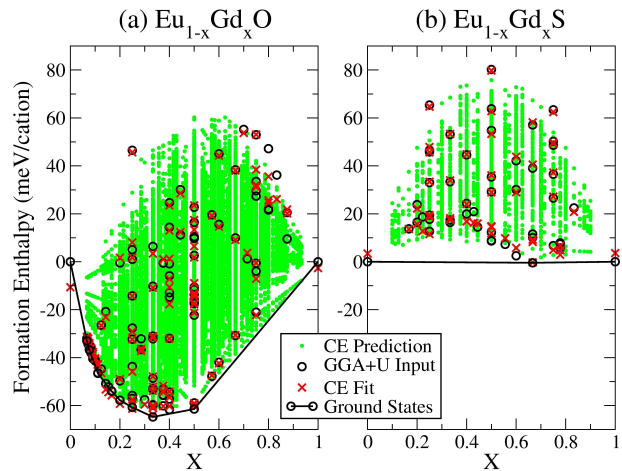}
\caption{(Color online) Formation enthalpy per cation {\emph{vs}} composition $x$
for all distinct cation orderings within rocksalt structure for
(a) Eu$_x$Gd$_{1-x}$O with up to 30 atoms per unit cell and (b)
Eu$_x$Gd$_{1-x}$S with up to 20 atoms per unit cell. The black open circles
are the first principles inputs, red crosses are the fitted CE values for the input structures,
and the green dots are the predicted $\Delta H_\mathrm{CE}$ for all other structures.
} \label{hulls}
\end{figure}

Figure~\ref{hulls}a shows the calculated formation enthalpy diagram for the
entire range of $x$ in Eu$_{1-x}$Gd$_{x}$O. Each green dot represents the
composition and the CE formation enthalpy $\Delta
H_{CE}(\boldsymbol{\sigma})$ of one of the $2^{15}$ structures
$\boldsymbol{\sigma}$ used in the ground state search. The fitted $\Delta
H_{CE}(\boldsymbol{\sigma}_\mathrm{in})$ values for the 148 input structures
are shown by the red crosses, and the input $\Delta
H_{GGA}(\boldsymbol{\sigma}_\mathrm{in})$  are shown by open black circles.
The input set contains eight stable compounds, serving as the breaking points
on the GGA+U convex hull. (This convex hull is shown by the black solid line
in Fig.\ \ref{hulls}.) These compounds represent tentative ground state
predictions, in a sense that they all were predicted as ground states
throughout the CE iterations and confirmed by direct GGA+U calculations. Most
of them have small energetic depths (see Table \ref{enthalpies}); the phase
diagram calculations show that only two of these, with $x=0.5$ and $x=1/3$,
can appear at reasonably high temperatures. Therefore, it is not necessary to
insist on the precise prediction of the ground state sequence at $x<1/3$.
Indeed, we found that the convex hull corresponding to our final CE
fitted Hamiltonian was not the same as that for the input set (i.e., it
changes even throughout the final CE iterations). However, full
self-consistency {\em was} achieved for $x\gtrsim 0.3$, allowing us to
identify the compounds with $x=1/3$ and 0.5 as unambiguously established
ground states. The structural information for these two compounds is given in
Appendix \ref{appendix:GSs}. Despite the slight disagreement regarding the identity of the
low-temperature ground states at $x<0.3$, the overall convex hull shape and
the $\Delta H_{CE}$ values agreed well between the input (GGA+U) and the
final predicted (CE) convex hulls, and the identity, formation enthalpies and
energetic depths of the $x=1/3$ and 0.5 ground states are accurately
reproduced.
Note also that the resulting CE fitting does not have an exact meaning of a formation
enthalpy, because the vanishing of the fitted quantity for the end members is
not enforced. In particular, pure EuO has a spurious ``formation enthalpy''
of $-10.6$ meV/cation in the CE fitting. However, the shape of the
ground-state convex hull suggests that the fit error for pure EuO should not
affect the phase diagram significantly at $x\gtrsim0.1$ where the fitting is
quite accurate.

\begin{table*}
\caption{Formation energies and the energetic depths
of the unambiguously established (regular font) and tentatively predicted ({\em italic})
ground-state structures for Eu$_{1-x}$Gd$_{x}$O and Eu$_{1-x}$Gd$_{x}$S systems.
A space
group notation or a \emph{Strukturbericht} designation of the cation
ordering (if available) is given in parenthesis after  the unit cell formula
of each compound.
$\Delta H_\mathrm{GGA}$, $\Delta
H_\mathrm{fixc}$, and $\Delta H_\mathrm{CE}$ are the formation enthalpies
obtained, respectively, from scalar-relativistic GGA+U with full relaxation,
from GGA+U with a restricted relaxation in which the cations are fixed at the undistorted fcc lattice with the Vegard-law lattice parameter (while the anions are allowed to relax), and from the CE fit. $\delta_\mathrm{GGA}$ and $\delta_\mathrm{CE}$ are the energetic
depths from the full relaxation and from the CE fit, respectively.
$\Delta H_\mathrm{GGA}^\mathrm{SO}$ are similar to $\Delta H_\mathrm{GGA}$
except the spin-orbit coupling was included.
All energetic quantities are given in
meV/cation.
}
\begin{ruledtabular}
\em
\begin{tabular}{lccrrrrr}
\em Formula unit & $x$ &
$\Delta H_\mathrm{GGA}$ \em ($\Delta H_\mathrm{GGA}^\mathrm{SO}$)
   & $\Delta H_\mathrm{fixc} $ & $\Delta H_\mathrm{CE}$ & $\delta_\mathrm{GGA}$ & $\delta_\mathrm{CE}$ \\
\hline
\multicolumn{7}{c}{ \vspace{0.03in} \em Eu$_{1-x}$Gd$_{x}$O: \vspace{2pt}}\\
\em
EuO (B1)                              & \em 0    & \em  0.0 &  \em  0.0   & \em -10.6                & &              \\
Eu$_{14}$GdO$_{15}$ (C2/m)            & 1/15 & -33.0 &  -32.8   & -31.2                & 1.1 & 0.5    \\
Eu$_{12}$GdO$_{13}$ (P$\overline1$)   & 1/13 & -36.8 &  -36.5   & -33.9                & 0.7 & 0.0    \\
Eu$_8$GdO$_9$ (P$\overline3$1m)       & 1/9  & -46.5 &  -46.4   & -42.6                & 3.2 & 0.4    \\
Eu$_{10}$Gd$_2$O$_{12}$ (C2/c)        & 1/6  & -54.0 &  -53.8   & -55.7                & 0.5 & 2.7    \\
Eu$_8$Gd$_2$O$_{10}$ (P$\overline1$)  & 1/5  & -57.8 &  -56.8   & -59.2                & 1.0 & 1.3    \\
Eu$_6$Gd$_2$O$_8$ (C2/c)              & 1/4  & -60.7 &  -58.3   & -61.3                & 0.4 & 0.4        \\
\em
Eu$_8$Gd$_4$O$_{12}$ (C2/m)           & \em  1/3  & \em  -64.7 (-62.8)& \em   -61.3   & \em  -63.7                & \em  3.8 & \em  2.9    \\
\em
EuGdO$_2$ (L1$_1$)                    & \em  1/2  & \em  -61.5 (-59.9)& \em   -60.6   & \em  -60.1                & \em 12.9 & \em 11.6    \\
\em
GdO (B1)                              & \em  1    & \em    0.0 & \em     0.0   & \em   -2.6                &      &        \\
\hline
\multicolumn{7}{c}{ \em Eu$_{1-x}$Gd$_{x}$S:  \vspace{2pt} }\\
\em
EuS (B1)                              & \em  0    & \em    0.0 & \em     0.0   & \em    3.4                &        &         \\
EuGd$_{2}$S$_{3}$ (C6)                & 2/3  &  -0.4 (-2.0)&   26.8   &  -0.3                & 0.4 & 3.8    \\
\em
GdS (B1)                              & \em  0    & \em    0.0 & \em     0.0   & \em    3.6                &      &         \\
\end{tabular}
\end{ruledtabular}
\label{enthalpies}
\end{table*}

The calculated formation enthalpies for both the unambiguously established and
tentatively predicted ground-state
Eu$_{1-x}$Gd$_{x}$O compounds identified
throughout the CE iterations and confirmed by direct GGA+U
calculations are listed in
Table~\ref{enthalpies} along with their CE-fitted values, as well as
the energetic depths $\delta$ found from both calculated and fitted formation
enthalpies.
The fit error
for all these compounds is less than 4 meV/cation.
The ground state with the largest $\delta$ has L1$_1$ structure, which is an
A$_1$B$_1$ (111) superlattice, i.e. it is formed by alternating (111) layers
of pure Eu and Gd. Among the six compounds with $x<1/3$ there are three other
pure Eu/Gd superlattices: A$_{14}$B (134), A$_{12}$B (124), and A$_8$B$_2$
(123), where the numbers in brackets denote the orientation of the pure
cation layers, A stands for Eu, and B for Gd. It is clear that no particular
superlattice direction is preferred.

For the region above $x=0.5$, our phase diagram predicts phase separation
into the L1$_1$ phase and pure GdO. Since GdO, as has been mentioned above,
has not been observed in the rocksalt structure considered here, we have
further investigated its stability. In the zincblende structure its energy is
found to be 56 meV/cation higher compared to rocksalt, but the equilibrium
lattice parameter 5.31 \AA~ is close to that ($5.24 \pm 0.05$ \AA) reported
\cite{Bist} for zincblende GdO. The fact that experimental data suggest
zincblende structure, while our calculation predicts it to be significantly
less stable than rocksalt, may be due to the likely off-stoichiometry in
experiment. Further, we considered the stability of GdO with respect to
decomposition into Gd$_2$O$_3$ and metallic Gd. Gd$_2$O$_3$ occurs in three
different structures \cite{sesquioxide}, cubic (80-atom unit cell) under 777
K, monoclinic (30-atom unit cell) from 777 to 1727 K, and hexagonal above
1727 K. We used fully relaxed lattice constants of Gd$_2$O$_3$, Gd and GdO.
Among the three phases of Gd$_2$O$_3$, the cubic phase is the most stable,
its energy per formula unit being lower by approximately 90 meV/cation
compared to the monoclinic phase. Rocksalt GdO energy was found to be 21
meV/cation (per Gd atom) lower compared to the mixture of Gd$_2$O$_3$ and Gd.
From this we can deduce that if the system is in equilibrium with an oxygen
reservoir, GdO can only form in a narrow range of the oxygen chemical
potential $\mu_\mathrm{O}$. Indeed, using the following relations:
\begin{equation}
\begin{split}
E_\mathrm{GdO}-E_\mathrm{Gd} < \mu_\mathrm{O} < E_\mathrm{Gd_2O_3}-2E_\mathrm{GdO}\\
E_\mathrm{Gd}+E_\mathrm{Gd_2O_3}=3E_\mathrm{GdO} + 63~\mathrm{meV}
\end{split}
\label{chempot}
\end{equation}
where the inequalities follow from the stability of GdO with respect to both
reduction to Gd and oxidation to Gd$_2$O$_3$, we find that the double
inequality is satisfied only in the 63 meV-wide range of $\mu_\mathrm{O}$. In
reality this range depends on temperature and is subject to the uncertainty
in the calculated reaction enthalpy, but nevertheless this feature is
consistent with the difficulty in stabilizing GdO experimentally.

\begin{figure}[!htb]
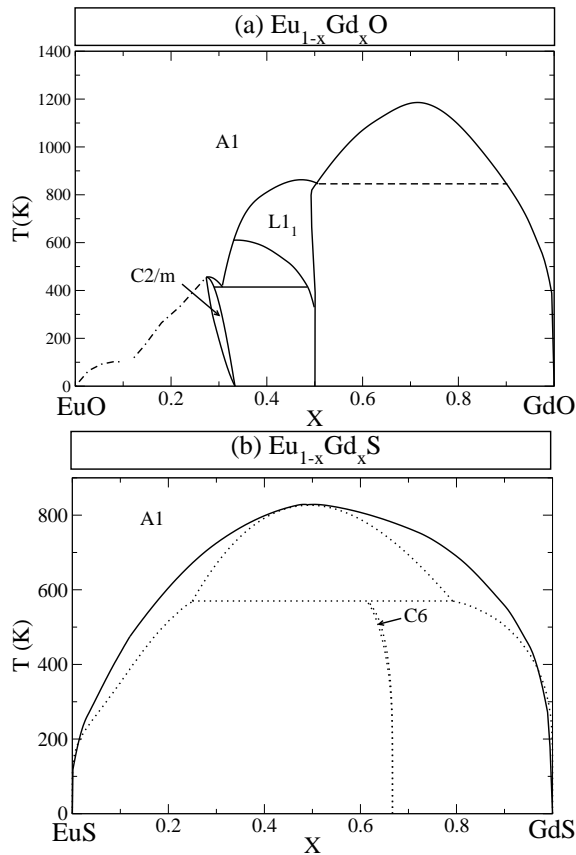

\centering
\begin{minipage}{8.5cm}
\begin{minipage}{8.5cm}
\includegraphics[width=7.5cm,angle=0,clip]{fig5a.eps}
\end{minipage}
\begin{minipage}{8.5cm}
\includegraphics[width=7.5cm,angle=0,clip]{fig5b.eps}
\end{minipage}
\end{minipage}
\caption{Calculated phase diagrams obtained using Monte Carlo simulations
from the cluster expansions for (a) Eu$_x$Gd$_{1-x}$O and (b)
Eu$_x$Gd$_{1-x}$S alloys. The phase labels indicate the ordering within the
cation sublattice. In (a), the phases in the region bounded by the dot-dashed
line have not been identified. In (b), the solid lines correspond to the
paramagnetic high-temperature phases, while the dotted lines show the
prediction of the $T=0$ ferromagnetic cluster expansion. } \label{PDs}
\end{figure}

The calculated phase diagram for the oxide system is shown in
Fig.~\ref{PDs}a. It is seen from Table~\ref{enthalpies} that most of the
ground states with $x<0.3$ have rather small energetic depths $\delta$,
suggesting that these orderings would only occur at very low temperatures.
Indeed, our simulations show that many of these phases only appear well below
room temperature, so that the corresponding phase transformations are
kinetically inaccessible. In fact, phase transformations occurring above
$T\sim400$ K involve only the phases with $x=1/3$ and $x=1/2$. Therefore, for
$x<0.3$ we have only determined the approximate boundary (i.\ e.\ the
solubility limit) of the disordered (Gd,Eu)O phase, which is shown by a
dot-dashed line in Fig.~\ref{PDs}a.

There are several interesting features in this phase diagram. (1) A broad
miscibility gap exists in the Gd-rich region with the critical point close to
1200 K and $x\approx0.7$. (2) A continuous order-disorder transition occurs
for the L1$_1$ phase, whose line terminates at a tricritical point ($T\approx
600$ K, $x\approx0.33$) on the Eu-rich end and at a critical endpoint at the
Gd-rich end ($T\approx850$ K, $x\approx0.51$). Thermodynamics mandates that
the slope of the solubility line at the tricritical point should be different
from the slope of the ordering line, but this difference is too small to be
revealed in Monte Carlo simulations. On the other hand, the slopes of the
binodals do not change at the critical endpoint, but their curvatures do. (3)
There is a eutectoid triple point at $T\approx 420$ K at which the disordered
phase decomposes in a mixture of C2/m and L1$_1$ phases. (4) A point of equal
concentrations caps the single-phase C2/m region; this point is close to the
eutectoid point.

Apart from the variations due to different phase transitions, the solubility
of GdO in EuO grows approximately linear with temperature up to the critical
point near 1200 K, with a slope of about 0.06\%/K.

As discussed in Appendix \ref{appendix:L11}, the ordering transition for the
prominent L1$_1$ ground state occurs at  $T_\mathrm{ord}=840$ K and is second
order around the stoichiometric EuGdO$_2$ composition. The physics behind
this ordering transition can be further illustrated by considering the CE
Hamiltonian [Eq.~(\ref{eq1})] as a generalized Ising model, in which Eu atoms
are represented by pseudo-spin $S=1/2$ and Gd's by $S=-1/2$, and the
pseudo-spins form an fcc lattice. In the L1$_1$ structure, the nearest
neighbor interaction is fully frustrated, while the second nearest neighbor
interaction is not frustrated (the second-nearest cation neighbors are always
of dislike type). Thus, we can regard this structure as being formed by four
interpenetrating simple cubic lattices, all of which have AFM ordered
pseudo-spins, and which are coupled only through longer-range interactions
and through the order-from-disorder mechanism. Indeed, the transition
temperature in our MC simulations (about $T_\mathrm{ord}=840$ K) is close to
that of the simple cubic lattice under the assumption that its first
nearest-neighbor interaction is equal to the second nearest neighbor
interaction of the original fcc lattice (920 K using the best known estimate
of $T_c$ in the AFM Ising model from Ref.\ \onlinecite{Ising}).

\subsection{Eu$_{1-x}$Gd$_{x}$S}
\label{sec:EuGdSstability}

The formation enthalpy diagram for the sulfide system (all compounds
including GdS are assumed to be FM) is shown in Fig.~\ref{hulls}b. This
sulfide system has a very different ground-state convex hull, compared to the
oxide system. There is only one marginally stable compound at $x=2/3$ with
the formation enthalpy of only $-0.4$ meV/cation (see
Table~\ref{enthalpies}); this number is, in fact, smaller than the precision
of our GGA+U calculations. This compound has the hexagonal C6 structure
(sometimes referenced as ``$\alpha2$'' in CE investigations); its structural
parameters are given in Appendix \ref{appendix:GSs}. This structure is a
superlattice composed of pure GdO and EuO (111) layers alternating in a 2:1
pattern (similar in this respect to the L1$_1$ structure which has a 1:1
pattern).

The computed phase diagram is depicted in Fig.~\ref{PDs}b. Dotted lines
correspond to the FM CE, and solid lines to the corrected CE, designed to
represent the PM phase equilibrium. This corrected CE was obtained by
changing the nearest-neighbor pair ECI in our FM CE from $-6.46$ meV to
$-7.37$ meV, which results in the correct paramagnetic formation enthalpy of
C6 EuGd$_2$S$_3$ (see Section \ref{sec:Magnetism} and Appendix
\ref{appendix:paramagnetic}).

In the FM phase diagram, there are two major features: a wide miscibility gap
with a critical point at $T\approx810$ K and $x\approx0.5$, and a peritectoid
triple point at $T\approx560$ K, at which the two disordered phases are in
equilibrium with a new predicted C6 phase. The C6 phase forms a very narrow
single-phase region with concentration slightly decreasing at elevated
temperatures. The very small energetic depth $\delta$ of the C6 phase is
significantly overestimated by the CE fit (see Table \ref{enthalpies}); we
therefore expect that the temperature of the peritectoid point is also
overestimated. In fact, in the paramagnetic phase diagram (solid lines) the
C6 phase is unstable ($\delta<0$). Note, however, that the GGA+U value of
$\delta$ (positive in the FM case and negative in the paramagnetic case) is
comparable to the precision of our GGA+U calculations; we thus conclude that
our accuracy is not sufficient to confidently select among the two scenarios
shown in Fig~\ref{PDs}b by the solid and dotted lines.

We have also calculated the enthalpy of the reaction similar to the second
line of Eq.\ \ref{chempot} with the $\alpha$-phase of Gd$_2$S$_3$ and
rocksalt GdS, which comes out at 2.8 eV. Thus, in equilibrium with a sulfur
reservoir the rocksalt GdS is stable against both reduction and oxidation in
the 2.8 eV-wide range of the S chemical potential. This is consistent
with the fact that this phase is readily obtained experimentally.

The overall shape of the sulfide phase diagram is much more symmetric
compared to the oxide system, which is consistent with weaker triplet ECI's
(see Fig.\ \ref{ECIplot}). However, the solubility limit of GdS in EuS is
still greater than that of EuS in GdS, similar to the oxide system. The
solubility limit of GdS in EuS is about 7\% at 300 K and increases almost
linearly at a rate of about 0.07\%/K.

\section{Anion-mediated deformational interaction}\label{sec:mechanism}

The ground states with L1$_1$ or a closely related D4 cation ordering (the
latter having the same number of like and dislike neighbors as the L1$_1$ for
{\em any} given neighbor distance), appear quite common for rocksalt
chalcogenides: in addition to EuGdO$_2$, such ground states were recently
predicted for a number of rocksalt tellurides.\cite{LAST,BAST} In the case of
tellurides, D4 appears a more typical ground state, although L1$_1$ typically
differs by only a few meV/cation in energy. Conversely, in EuGdO$_2$, D4 is
only 2.6 meV/cation higher in energy than L1$_1$ (which is sufficient for
L1$_1$ to prevail up to $T_\mathrm{ord}=840$ K, due to the near identity of
the entropic contributions to the free energy of both structures.) The
appearance of L1$_1$ and D4 ground states has been related to the elastic
softness to a deformation along the [111] direction,\cite{strainLAST} which
is indeed typical for rocksalt compounds, generally hardest along the [100]
cation-anion bond direction. Further, it appears that the other
high-$T_\mathrm{ord}$ phase that we have discussed, namely the tentatively
predicted C6 EuGd$_2$S$_3$ phase, follows the same elastic trend, since it is
a (111) superlattice just like L1$_1$.

It may not be immediately clear, however, why L1$_1$ EuGdS$_2$ is not
stable. More generally, while similar electronic structures for both oxides and sulfides are expected
from the isovalent electronic configurations (and confirmed by
first-principles calculations), their ground-state formation enthalpy
diagrams are very different: the oxides exhibit a convex hull with large and
negative formation enthalpies, but the sulfides have positive formation
enthalpies for all ordered compounds except one at $x=2/3$ which is close to
zero (Fig.~\ref{hulls} and Table~\ref{enthalpies}). We shall now
demonstrate that it is possible to understand the origin
of this difference, as well as to get a deeper understanding of the elastic
mechanism leading to the predominance of L1$_1$ and D4 in rocksalt
chalcogenide alloys, by examining the mechanism of atomic relaxation in these alloys.

Let us decompose the formation enthalpy in two parts: the
``unrelaxed chemical'' part $\Delta H_\mathrm{chem}$ which is calculated for the undistorted (cubic) lattice at the lattice parameter $a_{VL}$ given by the Vegard law,
and $\Delta H_{rel}$
due to the additional full relaxation from these ideal Vegard positions
(thus, $\Delta H=\Delta H_\mathrm{chem}+\Delta H_{rel}$).
This decomposition is shown in
Table~\ref{relaxation} for the L1$_0$ and L1$_1$ structures, which straddle
almost the entire range of formation enthalpies at $x=0.5$ for both oxide and
sulfide systems. The difference $\Delta H_{ord}$ between the enthalpies of
the L1$_0$ and L1$_1$ structures gives the characteristic ordering enthalpy;
its decomposition in the unrelaxed chemical and relaxation parts is also included
in Table~\ref{relaxation}.

\begin{table}
\caption{Decomposition of the formation enthalpy for L1$_0$ and L1$_1$
structures
in the unrelaxed chemical ($\Delta H_\mathrm{chem}$)
and relaxation ($\Delta H_{rel}$)
contributions (see text). $\Delta H^\mathrm{chem}_{ord}$ and $\Delta H^{rel}_{ord}$
are the corresponding differences between the L1$_0$ and L1$_1$ structures.
All enthalpies are given in meV/cation.}
\begin{ruledtabular}
\begin{tabular}{ccrrrrrr}
Composition & Structure & $\Delta H_\mathrm{chem}$ & $\Delta H_{rel}$ & $\Delta H^\mathrm{chem}_{ord}$ & $\Delta H^{rel}_{ord}$\\
\hline
\multirow{2}{*}{EuGdO$_2$}   & L1$_0$ & 63.1   &   -2.4   & \multirow{2}{*}{26.0} & \multirow{2}{*}{94.1} \\
                            & L1$_1$ & 37.1   &  -96.5   & &\\
\hline
\multirow{2}{*}{EuGdS$_2$}   & L1$_0$ & 88.1   &   -8.0   &  \multirow{2}{*}{25.5} & \multirow{2}{*}{45.0} \\
                            & L1$_1$ & 62.6   &  -53.0   & & \\
\end{tabular}

\end{ruledtabular}
\label{relaxation}
\end{table}

First, we see that the ``chemical'' part of the ordering energy $\Delta
H^\mathrm{chem}_{ord}$ is almost identical for both oxide and sulfide compounds,
reflecting the fact that they are isoelectronic and their bonding properties
are therefore very similar. Second, the values of $\Delta H_\mathrm{chem}$ are
significantly larger for both sulfide compounds compared to the oxide ones.
This feature can be explained by a notably larger lattice mismatch in the
sulfide system (see Table~\ref{elastic}), which forces the Gd-S and Eu-S
bonds to deviate further away from their equilibrium lengths.

For the tetragonal L1$_0$ cation ordering, symmetry forbids the relaxation of
all internal coordinates; only the lattice parameters are allowed to relax.
It is seen from Table~\ref{relaxation} that this relaxation yields only a
relatively small energy gain. However, in the rhombohedral L1$_1$ structure,
the coordinates of the oxygen atoms are not fixed by symmetry. The relaxation
of these coordinates makes the dominant contribution to $\Delta H_{rel}$,
which is almost twice as large in the oxide system as it is in the sulfide
system. Moreover, for L1$_1$ GdEuO$_2$ this internal relaxation overcomes the
positive $\Delta H_\mathrm{chem}$ contribution and makes the formation
enthalpy large and negative. In the sulfide system $\Delta H_{rel}$ is about
two times smaller and does not fully overcome the $\Delta H_\mathrm{chem}$
term.

Thus, we conclude that the L1$_1$ structure is strongly favored with respect
to L1$_0$ because it allows the anion atoms to relax. Specifically, consider
the octahedral cage occupied by an anion atom. In the L1$_0$ structure the Gd
and Eu layers are stacked along the [001] direction; both inequivalent anion
sites (in the Gd and Eu layers) are located at inversion centers.
Thus, the
opposite vertices of the octahedral cage are occupied by like cations (either
both Eu or both Gd). On the other hand, in the L1$_1$ phase
(or in the closely competing D4 phase) this inversion
symmetry is \emph{fully} broken in the sense that \emph{all} the opposing
vertices of all octahedral cages are occupied by unlike cations.
Since Gd and Eu ions have notably different radii, the anion atoms shift
towards Gd. The structure with the largest possible degree of inversion
symmetry breaking maximizes the ability of the system to gain energy from
anion relaxations. The opposite vertices of the octahedral cage are separated
by the next-nearest neighbor distance in the cation sublattice. Thus, the
relaxation of the anion atom connecting the next-nearest neighbor pair is
the mechanism generating the dominant positive ECI for this pair (see
Fig.~\ref{ECIplot}). The fact that this ECI is smaller for sulfides is fully
consistent with the data in Table~\ref{relaxation}, and is not surprising
because this anion-mediated interaction mechanism is sensitive to the anion
size. Indeed, one can argue that the problem of finding the ground state with
such interaction is largely equivalent to finding the most favorable packing
of atoms to minimize the deviations from the optimal bond lengths.

In order to understand the relative importance of the anion and cation
relaxations, we have recalculated the formation enthalpies of all compounds
by restricting the atomic relaxations to anions only and fixing the cations
to the sites of an ideal fcc lattice with the volume given by the Vegard law.
These values are listed in Table~\ref{enthalpies} as $\Delta
H_\mathrm{fixc}$. For all oxide compounds the formation enthalpies obtained
in this way agree very well with the fully relaxed formation enthalpies
($\Delta H$), indicating that anion relaxation in these structures are
unrestricted by symmetry whereas the cation relaxations are insignificant.
For the EuGd$_2$S$_3$ compound in the sulfide system, however, the
corresponding error is large. The reason is that this compound is a 2:1
layered superlattice, which makes it possible to adjust the interlayer Gd-S
and Eu-S bond lengths by changing the cation layer separations. This is not
possible in the L1$_1$ structure, which is a 1:1 layered superlattice,
because all the cation layer separations are equal there.

\section{Summary}\label{sec:summary}
We have computed the temperature-phase diagram of two isovalent alloy systems
Eu$_{1-x}$Gd$_x$O, Eu$_{1-x}$Gd$_{x}$S by using first-principles calculations
combined with the standard cluster expansion and Monte-Carlo simulations.
Very different ground-state convex hulls are obtained for the two systems:
the oxides form ordered compounds with large and negative formation
enthalpies, but sulfides have only one marginally stable compound. The
dominant configurational cation interaction comes from the second-nearest
neighbor pair and is mediated by the relaxation of the O atom lying in
between. The difference between oxides and sulfides is attributed to the
difference in the anion size. Gd has a high equilibrium solubility in EuO and
EuS even at room temperature, which indicates that rather heavy doping is
possible without the precipitation of secondary phases.

\begin{acknowledgments}
K.D.B. is grateful to Peter Dowben for numerous stimulating discussions. The
work at UNL was supported by the Defence Threat Reduction Agency (Grant No.
HDTRA1-07-1-0008) and the Nebraska Research Initiative. K.D.B. is a Cottrell
Scholar of Research Corporation. The work at UCLA was supported by the U.S.
Department of Energy, Nuclear Energy Research Initiative Consortium NERI-C
under Grant No. DE-FG07-07ID14893. The calculations performed at Oak Ridge
National Laboratory Center for Nanophase Materials Sciences were sponsored
by the Scientific User Facilities Division, Office of Basic Energy Sciences,
U.S. Department of Energy. The calculations have also been supported by the
National Science Foundation through TeraGrid resources\cite{TeraGrid}
provided by SDSC, NCSA, and Purdue University.
\end{acknowledgments}

\appendix

\section{Paramagnetic energies of sulfide compounds}
\label{appendix:paramagnetic} For pure EuS, GdS, and C6 EuGd$_2$S$_3$
structures predicted to be stable by the FM CE, we investigated the
paramagnetic phase regime by fitting magnetic configurational energies to the
classical Heisenberg model. For the pure GdS, up to the third-nearest-neigbor
exchange interactions were fitted using four magnetic configurations: (i) FM,
(ii) AFM type I (AFM$_\mathrm{I}$) along the [001] direction with alternating
spins, (iii) AFM type II (AFM$_\mathrm{II}$) along [111] with alternating
spins, and (iv) AFM type III (AFM$_\mathrm{III}$) along [001] with two layers
of alternating spins. As expected, we found that AFM$_\mathrm{II}$ is the
ground state for GdS with the mean-field transition temperature of 63 K
slightly above the experimental value 58 K. The fitting produces the
paramagnetic formation enthalpy of $-5.3$ meV relative to the FM
configuration.

The paramagnetic formation enthalpy of the pure EuS was also obtained using
the same magnetic configurations as for the pure GdS. The paramagnetic
formation enthalpy turns out to be 1.3 meV above the FM phase with the
mean-field value of the FM ordering temperature 10.2 K, somewhat less than
the experimental value of 19 K.

For the C6 EuGd$_2$S$_3$ structure, only the nearest-neigbor interactions
were considered but decomposed into four distinct types of Heisenberg
exchange parameters due to the layered structure and two cation species. In
the [111] direction, the C6 structure establishes an A$_2$B$_1$ type
superlattice, where each plane normal to the direction is composed of only
one cation species. This superlattice structure allows two types of
intra-layer exchange interactions, Gd-Gd and Eu-Eu, and two types of
inter-layer interactions, Gd-Gd and Gd-Eu. A total of nine magnetic
configurations were constructed in the 6-cation (1$\times$1$\times$2)
supercell doubled along the [111] direction of the 3-cation unit cell, and
also in the 6-cation (2$\times$1$\times$1) supercell doubled in the normal
direction to [111]. The magnetic configurations included, in the
1$\times$1$\times$2 supercell,
A$\uparrow$A$\uparrow$B$\uparrow$A$\uparrow$A$\uparrow$B$\uparrow$,
A$\uparrow$A$\downarrow$B$\uparrow$A$\downarrow$A$\uparrow$B$\downarrow$,
A$\downarrow$A$\uparrow$B$\downarrow$A$\downarrow$A$\uparrow$B$\uparrow$,
A$\downarrow$A$\downarrow$B$\uparrow$A$\uparrow$A$\uparrow$B$\downarrow$,
A$\uparrow$A$\downarrow$B$\downarrow$A$\downarrow$A$\uparrow$B$\uparrow$,
A$\uparrow$A$\uparrow$B$\uparrow$A$\downarrow$A$\uparrow$B$\downarrow$,
A$\uparrow$A$\uparrow$B$\downarrow$A$\uparrow$A$\uparrow$B$\downarrow$, and
in the 2$\times$1$\times$1 supercell, one of the two cations in either A or B
layer has its spin flipped, while all other cation spins are kept opposite in
directions. Its paramagnetic energy determined by the fitting with the four
exchange interaction parameters is slightly lower by 0.6 meV than the FM
configuration.

\section{Predicted stable compounds}
\label{appendix:GSs}
The predicted ground-state structures with reasonably high ordering
temperature are described below with their atomic positions fully relaxed
using VASP. Additionally, for each structure, its crystallographic space
group is identified for clarification.
\subsection{Eu$_8$Gd$_4$O$_{12}$}
\noindent Space group \#12 : C2/m, base-centered monoclinic  \\
Primitive unit-cell: a=12.507~\AA, b=7.187~\AA, c=8.766~\AA, \\
 $\alpha$=90.00$^\circ$, $\beta$=90.24$^\circ$, $\gamma$=90.00$^\circ$\\
Wyckoff positions:  Eu(2i)=(0.084,1/2,0.339),\\
Eu(2i)=(0.7503,1/2,0.995), Eu(4j)=(0.832,0.750,0.330), \\
Gd(2g)=(0,0.251,0), Gd(2i)=(0.417,1/2,0.667),\\
O(2i)=(0.075,1/2,0.851), O(4j)=(0.153,0.237.0.168),\\
O(2h)=(0,0.224,1/2), O(2i)=(0.260,1/2,0.512),\\
O(2i)=(0.432,1/2,0.159).
\subsection{EuGdO$_2$}
\noindent Space group \#166: R$\bar{3}$m, trigonal (rhombohedral)  \\
\textit{strukturbericht} designation of cation order: L1$_1$\\
Primitive unit-cell: a=3.536~\AA, b=3.536~\AA, c=17.741~\AA, \\
 $\alpha$=90.00$^\circ$, $\beta$=90.00$^\circ$, $\gamma$=120.00$^\circ$\\
Wyckoff positions: Eu(1b)=(0,0,1/2),\\
Gd(1a)=(0,0,0), O(2c)=(0,0,0.260).\\

\subsection{EuGd$_2$S$_3$}
\noindent Space group \#164: P$\bar{3}$m1, hexagonal  \\
\textit{strukturbericht} designation of cation order: C6\\
Primitive unit-cell: a=4.041~\AA, b=4.041~\AA, c=9.959~\AA, \\
 $\alpha$=90.00$^\circ$, $\beta$=90.00$^\circ$, $\gamma$=120.00$^\circ$\\
Wyckoff positions: Eu(1b)=(0,0,1/2),\\
Gd(2d)=(1/3,2/3,0.155), S(1a)=(0,0,0),\\
S(2d)=(1/3,2/3,0.681).

\section{Order-disorder transition in L1$_1$}
\label{appendix:L11}

\begin{figure}[b]
\includegraphics[width=7.5cm,angle=0,clip]{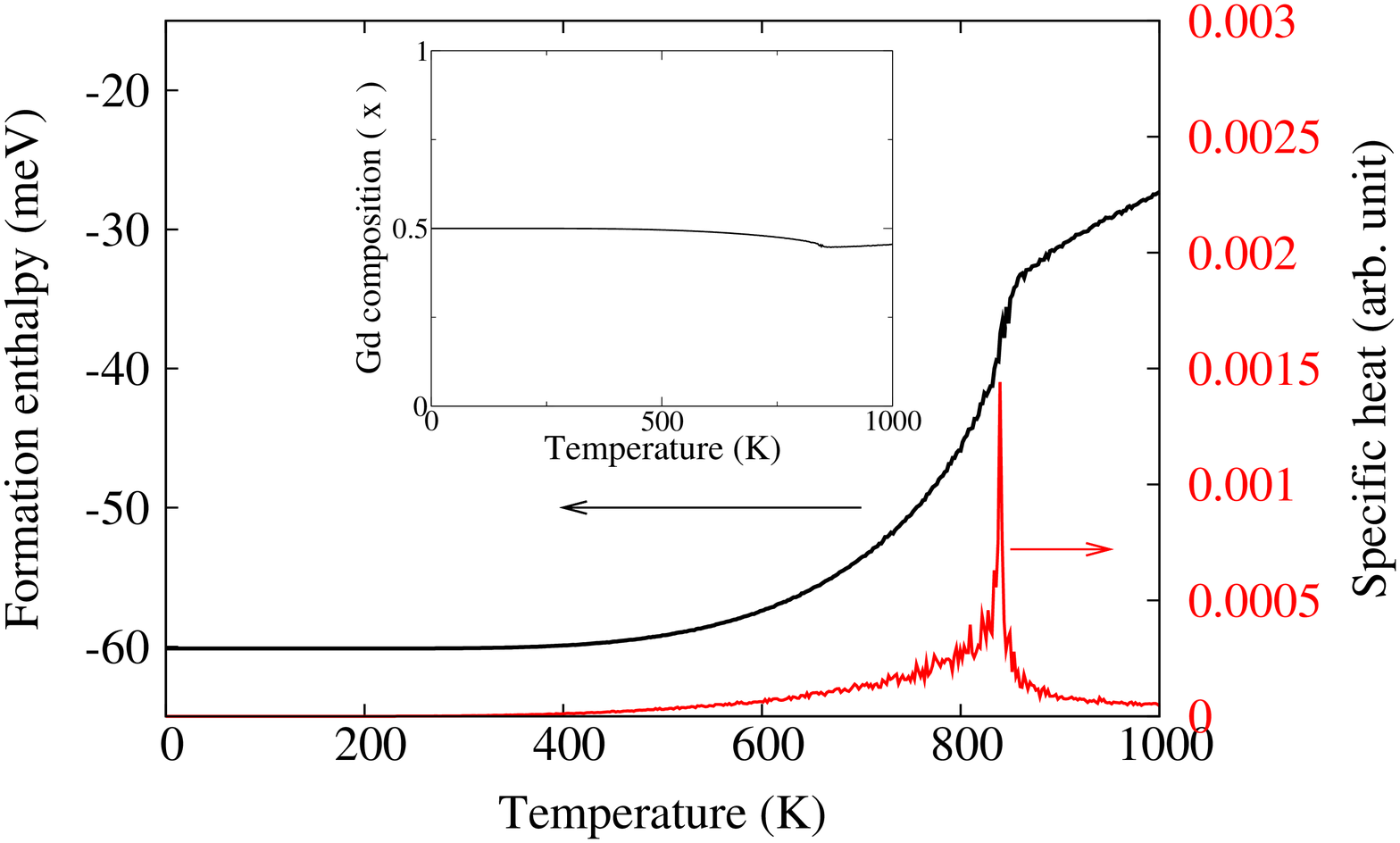}
\caption{(Color online) Monte Carlo heating simulation of the
L1$_1$-to-disordered phase transition for Eu$_{0.5}$Gd$_{0.5}$O. Formation
enthalpy and heat capacity are shown by black and red lines, respectively.
The inset shows the temperature dependence of the composition $x$. }
\label{L11OD}
\end{figure}

In finite-size MC simulations there is no formal distinction between a first-
and second-order transition, and the nature of the transition can be
unambiguously determined only from a finite-size scaling analysis. Although
we did not perform such an analysis, strong evidence in favor of the
second-order character of the ordering transition in L1$_1$ EuGdO$_2$ is
revealed by the behavior of the enthalpy and the heat capacity. A MC heating
simulation was performed starting with the L1$_1$ structure at $T=0$ K.
Figure~\ref{L11OD} shows the temperature dependence of the heat capacity, as
well as the total enthalpy, along the composition line corresponding to a
constant chemical potential $\mu=50$ meV (the concentration is shown in the
inset). The L1$_1$ phase persists up to 840 K, where a continuous
order-disorder transition is indicated by the characteristic peak of the heat
capacity, as well as the continuous change of the enthalpy.
Away from the Eu$_{0.5}$Gd$_{0.5}$O composition, the L1$_1$ ordered and the
disordered phases maintain the same features, although the heat capacity peak
is reduced.

\bibliography{paper}

\begin{thebibliography}{63}
\expandafter\ifx\csname natexlab\endcsname\relax\def\natexlab#1{#1}\fi
\expandafter\ifx\csname bibnamefont\endcsname\relax
  \def\bibnamefont#1{#1}\fi
\expandafter\ifx\csname bibfnamefont\endcsname\relax
  \def\bibfnamefont#1{#1}\fi
\expandafter\ifx\csname citenamefont\endcsname\relax
  \def\citenamefont#1{#1}\fi
\expandafter\ifx\csname url\endcsname\relax
  \def\url#1{\texttt{#1}}\fi
\expandafter\ifx\csname urlprefix\endcsname\relax\def\urlprefix{URL }\fi
\providecommand{\bibinfo}[2]{#2}
\providecommand{\eprint}[2][]{\url{#2}}

\bibitem[{\citenamefont{Samokhvalov et~al.}(1988)\citenamefont{Samokhvalov,
  Viglin, Gizhevskii, Arbuzova, and Chebotaev}}]{MIT}
\bibinfo{author}{\bibfnamefont{A.~A.} \bibnamefont{Samokhvalov}},
  \bibinfo{author}{\bibfnamefont{N.~A.} \bibnamefont{Viglin}},
  \bibinfo{author}{\bibfnamefont{B.~A.} \bibnamefont{Gizhevskii}},
  \bibinfo{author}{\bibfnamefont{T.~I.} \bibnamefont{Arbuzova}},
  \bibnamefont{and} \bibinfo{author}{\bibfnamefont{N.~M.}
  \bibnamefont{Chebotaev}}, \bibinfo{journal}{Phys. Status Solidi B}
  \textbf{\bibinfo{volume}{148}}, \bibinfo{pages}{361} (\bibinfo{year}{1988}).

\bibitem[{\citenamefont{Shapira et~al.}()\citenamefont{Shapira, Foner, and
  Reed}}]{MIT3}
\bibinfo{author}{\bibfnamefont{Y.}~\bibnamefont{Shapira}},
  \bibinfo{author}{\bibfnamefont{S.}~\bibnamefont{Foner}}, \bibnamefont{and}
  \bibinfo{author}{\bibfnamefont{T.~B.} \bibnamefont{Reed}},
  \bibinfo{note}{phys. Rev. B \textbf{8}, 2299 (1973); \textbf{8}, 2316
  (1973)}.

\bibitem[{\citenamefont{Steeneken et~al.}(2002)\citenamefont{Steeneken, Tjeng,
  Elfimov, Sawatzky, Ghiringhelli, Brookes, and Huang}}]{MIT2}
\bibinfo{author}{\bibfnamefont{P.~G.} \bibnamefont{Steeneken}},
  \bibinfo{author}{\bibfnamefont{L.~H.} \bibnamefont{Tjeng}},
  \bibinfo{author}{\bibfnamefont{I.}~\bibnamefont{Elfimov}},
  \bibinfo{author}{\bibfnamefont{G.~A.} \bibnamefont{Sawatzky}},
  \bibinfo{author}{\bibfnamefont{G.}~\bibnamefont{Ghiringhelli}},
  \bibinfo{author}{\bibfnamefont{N.~B.} \bibnamefont{Brookes}},
  \bibnamefont{and} \bibinfo{author}{\bibfnamefont{D.-J.} \bibnamefont{Huang}},
  \bibinfo{journal}{Phys. Rev. Lett.} \textbf{\bibinfo{volume}{88}},
  \bibinfo{pages}{047201} (\bibinfo{year}{2002}).

\bibitem[{\citenamefont{Schmehl et~al.}(2007)\citenamefont{Schmehl,
  Vaithyanathan, Herrnberger, Thiel, Richter, Liberati, Heeg, R\"ockerath,
  Kourkoutis, M\"uhlbauer et~al.}}]{eugdo3}
\bibinfo{author}{\bibfnamefont{A.}~\bibnamefont{Schmehl}},
  \bibinfo{author}{\bibfnamefont{V.}~\bibnamefont{Vaithyanathan}},
  \bibinfo{author}{\bibfnamefont{A.}~\bibnamefont{Herrnberger}},
  \bibinfo{author}{\bibfnamefont{S.}~\bibnamefont{Thiel}},
  \bibinfo{author}{\bibfnamefont{C.}~\bibnamefont{Richter}},
  \bibinfo{author}{\bibfnamefont{M.}~\bibnamefont{Liberati}},
  \bibinfo{author}{\bibfnamefont{T.}~\bibnamefont{Heeg}},
  \bibinfo{author}{\bibfnamefont{M.}~\bibnamefont{R\"ockerath}},
  \bibinfo{author}{\bibfnamefont{L.~F.} \bibnamefont{Kourkoutis}},
  \bibinfo{author}{\bibfnamefont{S.}~\bibnamefont{M\"uhlbauer}},
  \bibnamefont{et~al.}, \bibinfo{journal}{Nature Mater.}
  \textbf{\bibinfo{volume}{6}}, \bibinfo{pages}{882} (\bibinfo{year}{2007}).

\bibitem[{\citenamefont{Aoyama et~al.}(1992)\citenamefont{Aoyama, Oka, Honda,
  and Mori}}]{gd2}
\bibinfo{author}{\bibfnamefont{T.}~\bibnamefont{Aoyama}},
  \bibinfo{author}{\bibfnamefont{Y.}~\bibnamefont{Oka}},
  \bibinfo{author}{\bibfnamefont{K.}~\bibnamefont{Honda}}, \bibnamefont{and}
  \bibinfo{author}{\bibfnamefont{C.}~\bibnamefont{Mori}},
  \bibinfo{journal}{Nucl. Instr. Meth. Phys. Res. A}
  \textbf{\bibinfo{volume}{314}}, \bibinfo{pages}{590} (\bibinfo{year}{1992}).

\bibitem[{\citenamefont{Mireshghi et~al.}(1994)\citenamefont{Mireshghi, Cho,
  Drewery, Hong, Jing, Lee, Kaplan, and Perez-Mendez}}]{gd1}
\bibinfo{author}{\bibfnamefont{A.}~\bibnamefont{Mireshghi}},
  \bibinfo{author}{\bibfnamefont{G.}~\bibnamefont{Cho}},
  \bibinfo{author}{\bibfnamefont{J.~S.} \bibnamefont{Drewery}},
  \bibinfo{author}{\bibfnamefont{W.~S.} \bibnamefont{Hong}},
  \bibinfo{author}{\bibfnamefont{T.}~\bibnamefont{Jing}},
  \bibinfo{author}{\bibfnamefont{H.}~\bibnamefont{Lee}},
  \bibinfo{author}{\bibfnamefont{S.~N.} \bibnamefont{Kaplan}},
  \bibnamefont{and}
  \bibinfo{author}{\bibfnamefont{V.}~\bibnamefont{Perez-Mendez}},
  \bibinfo{journal}{IEEE Transactions on Nuclear Science}
  \textbf{\bibinfo{volume}{41}}, \bibinfo{pages}{915} (\bibinfo{year}{1994}).

\bibitem[{\citenamefont{Ketsman et~al.}(2007)\citenamefont{Ketsman, Losovyj,
  Sokolov, Tang, Wang, Belashchenko, and Dowben}}]{Ketsman}
\bibinfo{author}{\bibfnamefont{I.}~\bibnamefont{Ketsman}},
  \bibinfo{author}{\bibfnamefont{Y.~B.} \bibnamefont{Losovyj}},
  \bibinfo{author}{\bibfnamefont{A.}~\bibnamefont{Sokolov}},
  \bibinfo{author}{\bibfnamefont{J.}~\bibnamefont{Tang}},
  \bibinfo{author}{\bibfnamefont{Z.}~\bibnamefont{Wang}},
  \bibinfo{author}{\bibfnamefont{K.~D.} \bibnamefont{Belashchenko}},
  \bibnamefont{and} \bibinfo{author}{\bibfnamefont{P.~A.}
  \bibnamefont{Dowben}}, \bibinfo{journal}{Appl. Phys. A}
  \textbf{\bibinfo{volume}{89}}, \bibinfo{pages}{489} (\bibinfo{year}{2007}).

\bibitem[{\citenamefont{Schultz et~al.}(2010)\citenamefont{Schultz, Blasy,
  Santana, Young, Petrosky, McClory, LaGraffe, Brand, Tang, Wang
  et~al.}}]{Schultz}
\bibinfo{author}{\bibfnamefont{D.}~\bibnamefont{Schultz}},
  \bibinfo{author}{\bibfnamefont{B.}~\bibnamefont{Blasy}},
  \bibinfo{author}{\bibfnamefont{J.~C.} \bibnamefont{Santana}},
  \bibinfo{author}{\bibfnamefont{C.}~\bibnamefont{Young}},
  \bibinfo{author}{\bibfnamefont{J.~C.} \bibnamefont{Petrosky}},
  \bibinfo{author}{\bibfnamefont{J.~W.} \bibnamefont{McClory}},
  \bibinfo{author}{\bibfnamefont{D.}~\bibnamefont{LaGraffe}},
  \bibinfo{author}{\bibfnamefont{J.~I.} \bibnamefont{Brand}},
  \bibinfo{author}{\bibfnamefont{J.}~\bibnamefont{Tang}},
  \bibinfo{author}{\bibfnamefont{W.}~\bibnamefont{Wang}}, \bibnamefont{et~al.},
  \bibinfo{journal}{J. Phys. D: Appl. Phys.} \textbf{\bibinfo{volume}{43}},
  \bibinfo{pages}{075502} (\bibinfo{year}{2010}).

\bibitem[{\citenamefont{Eliseev et~al.}(1989)\citenamefont{Eliseev, Sadovskaya,
  and Pavlov}}]{eliseev89}
\bibinfo{author}{\bibfnamefont{A.~A.} \bibnamefont{Eliseev}},
  \bibinfo{author}{\bibfnamefont{O.~A.} \bibnamefont{Sadovskaya}},
  \bibnamefont{and} \bibinfo{author}{\bibfnamefont{A.~V.}
  \bibnamefont{Pavlov}}, \bibinfo{journal}{Inorg.Mater.}
  \textbf{\bibinfo{volume}{25}}, \bibinfo{pages}{10} (\bibinfo{year}{1989}).

\bibitem[{\citenamefont{Dimmock}(1970)}]{euo1}
\bibinfo{author}{\bibfnamefont{J.~O.} \bibnamefont{Dimmock}},
  \bibinfo{journal}{IBM J. Res. Dev.} \textbf{\bibinfo{volume}{314}},
  \bibinfo{pages}{301} (\bibinfo{year}{1970}).

\bibitem[{\citenamefont{Matsumoto et~al.}(2004)\citenamefont{Matsumoto,
  Yamaguchi, Yuri, Kawaguchi, Koshizaki, and Yamada}}]{mag1}
\bibinfo{author}{\bibfnamefont{T.}~\bibnamefont{Matsumoto}},
  \bibinfo{author}{\bibfnamefont{K.}~\bibnamefont{Yamaguchi}},
  \bibinfo{author}{\bibfnamefont{M.}~\bibnamefont{Yuri}},
  \bibinfo{author}{\bibfnamefont{K.}~\bibnamefont{Kawaguchi}},
  \bibinfo{author}{\bibfnamefont{N.}~\bibnamefont{Koshizaki}},
  \bibnamefont{and} \bibinfo{author}{\bibfnamefont{K.}~\bibnamefont{Yamada}},
  \bibinfo{journal}{J. Phys.: Condens. Matter} \textbf{\bibinfo{volume}{16}},
  \bibinfo{pages}{6017–6028} (\bibinfo{year}{2004}).

\bibitem[{\citenamefont{Ott et~al.}(2006)\citenamefont{Ott, Heise, Sutarto, Hu,
  Chang, Hsieh, Lin, Chen, and Tjeng}}]{mag2}
\bibinfo{author}{\bibfnamefont{H.}~\bibnamefont{Ott}},
  \bibinfo{author}{\bibfnamefont{S.~J.} \bibnamefont{Heise}},
  \bibinfo{author}{\bibfnamefont{R.}~\bibnamefont{Sutarto}},
  \bibinfo{author}{\bibfnamefont{Z.}~\bibnamefont{Hu}},
  \bibinfo{author}{\bibfnamefont{C.~F.} \bibnamefont{Chang}},
  \bibinfo{author}{\bibfnamefont{H.~H.} \bibnamefont{Hsieh}},
  \bibinfo{author}{\bibfnamefont{H.-J.} \bibnamefont{Lin}},
  \bibinfo{author}{\bibfnamefont{C.~T.} \bibnamefont{Chen}}, \bibnamefont{and}
  \bibinfo{author}{\bibfnamefont{L.~H.} \bibnamefont{Tjeng}},
  \bibinfo{journal}{Phys. Rev. B} \textbf{\bibinfo{volume}{73}},
  \bibinfo{pages}{094407} (\bibinfo{year}{2006}).

\bibitem[{\citenamefont{Shafer and McGuire}(1968)}]{eugdo1}
\bibinfo{author}{\bibfnamefont{M.~W.} \bibnamefont{Shafer}} \bibnamefont{and}
  \bibinfo{author}{\bibfnamefont{T.~R.} \bibnamefont{McGuire}},
  \bibinfo{journal}{J. Appl. Phys.} \textbf{\bibinfo{volume}{39}},
  \bibinfo{pages}{588} (\bibinfo{year}{1968}).

\bibitem[{\citenamefont{Mauger}(1977)}]{eugdo2}
\bibinfo{author}{\bibfnamefont{A.}~\bibnamefont{Mauger}},
  \bibinfo{journal}{Phys. Stat. Sol. (b)} \textbf{\bibinfo{volume}{84}},
  \bibinfo{pages}{761} (\bibinfo{year}{1977}).

\bibitem[{\citenamefont{Adachi and Imanaka}(1998)}]{sesquioxide}
\bibinfo{author}{\bibfnamefont{G.}~\bibnamefont{Adachi}} \bibnamefont{and}
  \bibinfo{author}{\bibfnamefont{N.}~\bibnamefont{Imanaka}},
  \bibinfo{journal}{Chem. Rev.} \textbf{\bibinfo{volume}{98}},
  \bibinfo{pages}{1479} (\bibinfo{year}{1998}).

\bibitem[{\citenamefont{Bist and Srivastava}(1971)}]{Bist}
\bibinfo{author}{\bibfnamefont{B.~M.~S.} \bibnamefont{Bist}} \bibnamefont{and}
  \bibinfo{author}{\bibfnamefont{O.~N.} \bibnamefont{Srivastava}},
  \bibinfo{journal}{Phys. Stat. Sol. (a)} \textbf{\bibinfo{volume}{7}},
  \bibinfo{pages}{K9} (\bibinfo{year}{1971}).

\bibitem[{\citenamefont{McGuire and Holtzberg}(1972)}]{eugds1}
\bibinfo{author}{\bibfnamefont{T.~R.} \bibnamefont{McGuire}} \bibnamefont{and}
  \bibinfo{author}{\bibfnamefont{F.}~\bibnamefont{Holtzberg}},
  \bibinfo{journal}{AIP Conf. Proc.} \textbf{\bibinfo{volume}{5}},
  \bibinfo{pages}{855} (\bibinfo{year}{1972}).

\bibitem[{\citenamefont{Kunes et~al.}(2005)\citenamefont{Kunes, Ku, and
  Pickett}}]{euo0}
\bibinfo{author}{\bibfnamefont{J.}~\bibnamefont{Kunes}},
  \bibinfo{author}{\bibfnamefont{W.}~\bibnamefont{Ku}}, \bibnamefont{and}
  \bibinfo{author}{\bibfnamefont{W.~P.} \bibnamefont{Pickett}},
  \bibinfo{journal}{J. Phys. Soc. Japan} \textbf{\bibinfo{volume}{74}},
  \bibinfo{pages}{1408} (\bibinfo{year}{2005}).

\bibitem[{\citenamefont{Larson and Lambrecht}(2006)}]{Larson}
\bibinfo{author}{\bibfnamefont{P.}~\bibnamefont{Larson}} \bibnamefont{and}
  \bibinfo{author}{\bibfnamefont{W.~R.~L.} \bibnamefont{Lambrecht}},
  \bibinfo{journal}{J. Phys.: Condens. Matter} \textbf{\bibinfo{volume}{18}},
  \bibinfo{pages}{11333} (\bibinfo{year}{2006}).

\bibitem[{\citenamefont{Ghosh et~al.}(2004)\citenamefont{Ghosh, De, and
  De}}]{GhoshEuO}
\bibinfo{author}{\bibfnamefont{D.~B.} \bibnamefont{Ghosh}},
  \bibinfo{author}{\bibfnamefont{M.}~\bibnamefont{De}}, \bibnamefont{and}
  \bibinfo{author}{\bibfnamefont{S.~K.} \bibnamefont{De}},
  \bibinfo{journal}{Phys. Rev. B} \textbf{\bibinfo{volume}{70}},
  \bibinfo{pages}{115211} (\bibinfo{year}{2004}).

\bibitem[{\citenamefont{Ingle and Elfimov}(2008)}]{Ingle08}
\bibinfo{author}{\bibfnamefont{N.~J.~C.} \bibnamefont{Ingle}} \bibnamefont{and}
  \bibinfo{author}{\bibfnamefont{I.~S.} \bibnamefont{Elfimov}},
  \bibinfo{journal}{Phys. Rev. B} \textbf{\bibinfo{volume}{77}},
  \bibinfo{pages}{121202(R)} (\bibinfo{year}{2008}).

\bibitem[{\citenamefont{Bousquet et~al.}(2010)\citenamefont{Bousquet, Spaldin,
  and Ghosez}}]{EuOferroel}
\bibinfo{author}{\bibfnamefont{E.}~\bibnamefont{Bousquet}},
  \bibinfo{author}{\bibfnamefont{N.~A.} \bibnamefont{Spaldin}},
  \bibnamefont{and} \bibinfo{author}{\bibfnamefont{P.}~\bibnamefont{Ghosez}},
  \bibinfo{journal}{Phys. Rev. Lett.} \textbf{\bibinfo{volume}{104}},
  \bibinfo{pages}{037601} (\bibinfo{year}{2010}).

\bibitem[{\citenamefont{Sinjukow and Nolting}(2003)}]{Nolting1}
\bibinfo{author}{\bibfnamefont{P.}~\bibnamefont{Sinjukow}} \bibnamefont{and}
  \bibinfo{author}{\bibfnamefont{W.}~\bibnamefont{Nolting}},
  \bibinfo{journal}{Phys. Rev. B} \textbf{\bibinfo{volume}{68}},
  \bibinfo{pages}{125107} (\bibinfo{year}{2003}).

\bibitem[{\citenamefont{Sinjukow and Nolting}(2004)}]{Nolting2}
\bibinfo{author}{\bibfnamefont{P.}~\bibnamefont{Sinjukow}} \bibnamefont{and}
  \bibinfo{author}{\bibfnamefont{W.}~\bibnamefont{Nolting}},
  \bibinfo{journal}{Phys. Rev. B} \textbf{\bibinfo{volume}{69}},
  \bibinfo{pages}{214432} (\bibinfo{year}{2004}).

\bibitem[{\citenamefont{Arnold and Kroha}(2008)}]{Arnold}
\bibinfo{author}{\bibfnamefont{M.}~\bibnamefont{Arnold}} \bibnamefont{and}
  \bibinfo{author}{\bibfnamefont{J.}~\bibnamefont{Kroha}},
  \bibinfo{journal}{Phys. Rev. Lett.} \textbf{\bibinfo{volume}{100}},
  \bibinfo{pages}{046404} (\bibinfo{year}{2008}).

\bibitem[{\citenamefont{Bl\"ochl}(1994)}]{Bloechl}
\bibinfo{author}{\bibfnamefont{P.~E.} \bibnamefont{Bl\"ochl}},
  \bibinfo{journal}{Phys. Rev. B} \textbf{\bibinfo{volume}{50}},
  \bibinfo{pages}{17953} (\bibinfo{year}{1994}).

\bibitem[{\citenamefont{Kresse and Joubert}(1999)}]{VASP-PAW}
\bibinfo{author}{\bibfnamefont{G.}~\bibnamefont{Kresse}} \bibnamefont{and}
  \bibinfo{author}{\bibfnamefont{D.}~\bibnamefont{Joubert}},
  \bibinfo{journal}{Phys. Rev. B} \textbf{\bibinfo{volume}{59}},
  \bibinfo{pages}{1758} (\bibinfo{year}{1999}).

\bibitem[{\citenamefont{Perdew et~al.}(1996)\citenamefont{Perdew, Burke, and
  Ernzerhof}}]{PBE}
\bibinfo{author}{\bibfnamefont{J.~P.} \bibnamefont{Perdew}},
  \bibinfo{author}{\bibfnamefont{K.}~\bibnamefont{Burke}}, \bibnamefont{and}
  \bibinfo{author}{\bibfnamefont{M.}~\bibnamefont{Ernzerhof}},
  \bibinfo{journal}{Phys. Rev. Lett.} \textbf{\bibinfo{volume}{77}},
  \bibinfo{pages}{3865} (\bibinfo{year}{1996}).

\bibitem[{\citenamefont{Liechtenstein et~al.}(1995)\citenamefont{Liechtenstein,
  Anisimov, and Zaanen}}]{ldau}
\bibinfo{author}{\bibfnamefont{A.~I.} \bibnamefont{Liechtenstein}},
  \bibinfo{author}{\bibfnamefont{V.~I.} \bibnamefont{Anisimov}},
  \bibnamefont{and} \bibinfo{author}{\bibfnamefont{J.}~\bibnamefont{Zaanen}},
  \bibinfo{journal}{Phys. Rev. B} \textbf{\bibinfo{volume}{52}},
  \bibinfo{pages}{R5467} (\bibinfo{year}{1995}).

\bibitem[{VAS()}]{VASP}
\bibinfo{note}{G. Kresse and J. Hafner, Phys. Rev. B \textbf{48}, 13115 (1993);
  G. Kresse and J. Furthm\"uller, Comput. Mater. Sci. \textbf{6}, 15 (1996);
  Phys. Rev. B \textbf{54}, 11169 (1996).}

\bibitem[{\citenamefont{Dederichs et~al.}(1984)\citenamefont{Dederichs,
  Bl\"ugel, Zeller, and Akai}}]{constr1}
\bibinfo{author}{\bibfnamefont{P.~H.} \bibnamefont{Dederichs}},
  \bibinfo{author}{\bibfnamefont{S.}~\bibnamefont{Bl\"ugel}},
  \bibinfo{author}{\bibfnamefont{R.}~\bibnamefont{Zeller}}, \bibnamefont{and}
  \bibinfo{author}{\bibfnamefont{H.}~\bibnamefont{Akai}},
  \bibinfo{journal}{Phys. Rev. Lett.} \textbf{\bibinfo{volume}{53}},
  \bibinfo{pages}{2512} (\bibinfo{year}{1984}).

\bibitem[{\citenamefont{Solovyev}(2008)}]{constr2}
\bibinfo{author}{\bibfnamefont{I.~V.} \bibnamefont{Solovyev}},
  \bibinfo{journal}{J. Phys.: Condens. Matter} \textbf{\bibinfo{volume}{20}},
  \bibinfo{pages}{293201} (\bibinfo{year}{2008}).

\bibitem[{FLE()}]{FLEUR}
\bibinfo{note}{\url{www.flapw.de}}.

\bibitem[{\citenamefont{Busch and Wachter}(1966)}]{euo3}
\bibinfo{author}{\bibfnamefont{G.}~\bibnamefont{Busch}} \bibnamefont{and}
  \bibinfo{author}{\bibfnamefont{P.}~\bibnamefont{Wachter}},
  \bibinfo{journal}{Phys. Kondens. Materie} \textbf{\bibinfo{volume}{5}},
  \bibinfo{pages}{232} (\bibinfo{year}{1966}).

\bibitem[{\citenamefont{Busch and Wachter}(1968)}]{euo4}
\bibinfo{author}{\bibfnamefont{G.}~\bibnamefont{Busch}} \bibnamefont{and}
  \bibinfo{author}{\bibfnamefont{P.}~\bibnamefont{Wachter}},
  \bibinfo{journal}{Z. Angew. Phys.} \textbf{\bibinfo{volume}{26}},
  \bibinfo{pages}{2} (\bibinfo{year}{1968}).

\bibitem[{\citenamefont{Eastman et~al.}(1969)\citenamefont{Eastman, Holtzberg,
  and Methfessel}}]{Eastman}
\bibinfo{author}{\bibfnamefont{D.~E.} \bibnamefont{Eastman}},
  \bibinfo{author}{\bibfnamefont{F.}~\bibnamefont{Holtzberg}},
  \bibnamefont{and}
  \bibinfo{author}{\bibfnamefont{S.}~\bibnamefont{Methfessel}},
  \bibinfo{journal}{Phys. Rev. Lett.} \textbf{\bibinfo{volume}{23}},
  \bibinfo{pages}{226} (\bibinfo{year}{1969}).

\bibitem[{\citenamefont{Zhou and V.Ozolins}(2009)}]{FeiSIC}
\bibinfo{author}{\bibfnamefont{F.}~\bibnamefont{Zhou}} \bibnamefont{and}
  \bibinfo{author}{\bibnamefont{V.Ozolins}}, \bibinfo{journal}{Phys. Rev. B}
  \textbf{\bibinfo{volume}{80}}, \bibinfo{pages}{125127}
  (\bibinfo{year}{2009}).

\bibitem[{\citenamefont{Wood and Zunger}(1985)}]{RMM}
\bibinfo{author}{\bibfnamefont{D.~M.} \bibnamefont{Wood}} \bibnamefont{and}
  \bibinfo{author}{\bibfnamefont{A.}~\bibnamefont{Zunger}},
  \bibinfo{journal}{J. Phys. A} \textbf{\bibinfo{volume}{18}},
  \bibinfo{pages}{1343} (\bibinfo{year}{1985}).

\bibitem[{\citenamefont{Davidson}(1983)}]{Davidson}
\bibinfo{author}{\bibfnamefont{E.}~\bibnamefont{Davidson}},
  \emph{\bibinfo{title}{Methods in Computational Molecular Physics, ed. by
  G.H.F. Diercksen and S. Wilson}} (\bibinfo{publisher}{Plenum},
  \bibinfo{address}{New York}, \bibinfo{year}{1983}), vol.
  \bibinfo{volume}{113} of \emph{\bibinfo{series}{NATO Advanced Study
  Institute, Series C}}, p.~\bibinfo{pages}{95}.

\bibitem[{\citenamefont{Kotani et~al.}(2007)\citenamefont{Kotani, van
  Schilfgaarde, and Faleev}}]{Kotani07}
\bibinfo{author}{\bibfnamefont{T.}~\bibnamefont{Kotani}},
  \bibinfo{author}{\bibfnamefont{M.}~\bibnamefont{van Schilfgaarde}},
  \bibnamefont{and} \bibinfo{author}{\bibfnamefont{S.~V.}
  \bibnamefont{Faleev}}, \bibinfo{journal}{Phys. Rev. B}
  \textbf{\bibinfo{volume}{76}}, \bibinfo{pages}{165106}
  (\bibinfo{year}{2007}).

\bibitem[{\citenamefont{Faleev et~al.}(2004)\citenamefont{Faleev, van
  Schilfgaarde, and Kotani}}]{Faleev04}
\bibinfo{author}{\bibfnamefont{S.~V.} \bibnamefont{Faleev}},
  \bibinfo{author}{\bibfnamefont{M.}~\bibnamefont{van Schilfgaarde}},
  \bibnamefont{and} \bibinfo{author}{\bibfnamefont{T.}~\bibnamefont{Kotani}},
  \bibinfo{journal}{Phys. Rev. Lett.} \textbf{\bibinfo{volume}{93}},
  \bibinfo{pages}{126406} (\bibinfo{year}{2004}).

\bibitem[{\citenamefont{van Schilfgaarde et~al.}(2006)\citenamefont{van
  Schilfgaarde, Kotani, and Faleev}}]{mark06qsgw}
\bibinfo{author}{\bibfnamefont{M.}~\bibnamefont{van Schilfgaarde}},
  \bibinfo{author}{\bibfnamefont{T.}~\bibnamefont{Kotani}}, \bibnamefont{and}
  \bibinfo{author}{\bibfnamefont{S.}~\bibnamefont{Faleev}},
  \bibinfo{journal}{Phys. Rev. Lett.} \textbf{\bibinfo{volume}{96}},
  \bibinfo{eid}{226402} (pages~\bibinfo{numpages}{4}) (\bibinfo{year}{2006}).

\bibitem[{\citenamefont{Chantis et~al.}(2006)\citenamefont{Chantis, van
  Schilfgaarde, and Kotani}}]{Chantis06a}
\bibinfo{author}{\bibfnamefont{A.~N.} \bibnamefont{Chantis}},
  \bibinfo{author}{\bibfnamefont{M.}~\bibnamefont{van Schilfgaarde}},
  \bibnamefont{and} \bibinfo{author}{\bibfnamefont{T.}~\bibnamefont{Kotani}},
  \bibinfo{journal}{Phys. Rev. Lett.} \textbf{\bibinfo{volume}{96}},
  \bibinfo{eid}{086405} (\bibinfo{year}{2006}).

\bibitem[{\citenamefont{Chantis et~al.}(2007)\citenamefont{Chantis, van
  Schilfgaarde, and Kotani}}]{Chantis07a}
\bibinfo{author}{\bibfnamefont{A.~N.} \bibnamefont{Chantis}},
  \bibinfo{author}{\bibfnamefont{M.}~\bibnamefont{van Schilfgaarde}},
  \bibnamefont{and} \bibinfo{author}{\bibfnamefont{T.}~\bibnamefont{Kotani}},
  \bibinfo{journal}{Phys. Rev. B} \textbf{\bibinfo{volume}{76}},
  \bibinfo{pages}{165126} (\bibinfo{year}{2007}).

\bibitem[{\citenamefont{Bruneval et~al.}(2006)\citenamefont{Bruneval, Vast,
  Reining, Izquierdo, Sirotti, and Barrett}}]{Bruneval06b}
\bibinfo{author}{\bibfnamefont{F.}~\bibnamefont{Bruneval}},
  \bibinfo{author}{\bibfnamefont{N.}~\bibnamefont{Vast}},
  \bibinfo{author}{\bibfnamefont{L.}~\bibnamefont{Reining}},
  \bibinfo{author}{\bibfnamefont{M.}~\bibnamefont{Izquierdo}},
  \bibinfo{author}{\bibfnamefont{F.}~\bibnamefont{Sirotti}}, \bibnamefont{and}
  \bibinfo{author}{\bibfnamefont{N.}~\bibnamefont{Barrett}},
  \bibinfo{journal}{Phys. Rev. Lett.} \textbf{\bibinfo{volume}{97}},
  \bibinfo{pages}{267601} (\bibinfo{year}{2006}).

\bibitem[{\citenamefont{Chambers et~al.}(2004)\citenamefont{Chambers, Droubay,
  Kaspar, Gutowski, and van Schilfgaarde}}]{Chambers04}
\bibinfo{author}{\bibfnamefont{S.}~\bibnamefont{Chambers}},
  \bibinfo{author}{\bibfnamefont{T.}~\bibnamefont{Droubay}},
  \bibinfo{author}{\bibfnamefont{T.}~\bibnamefont{Kaspar}},
  \bibinfo{author}{\bibfnamefont{M.}~\bibnamefont{Gutowski}}, \bibnamefont{and}
  \bibinfo{author}{\bibfnamefont{M.}~\bibnamefont{van Schilfgaarde}},
  \bibinfo{journal}{Surf. Sci. Lett.} \textbf{\bibinfo{volume}{554}},
  \bibinfo{pages}{81} (\bibinfo{year}{2004}).

\bibitem[{\citenamefont{Shishkin et~al.}(2007)\citenamefont{Shishkin, Marsman,
  and Kresse}}]{Shishkin07}
\bibinfo{author}{\bibfnamefont{M.}~\bibnamefont{Shishkin}},
  \bibinfo{author}{\bibfnamefont{M.}~\bibnamefont{Marsman}}, \bibnamefont{and}
  \bibinfo{author}{\bibfnamefont{G.}~\bibnamefont{Kresse}},
  \bibinfo{journal}{Phys. Rev. Lett.} \textbf{\bibinfo{volume}{99}},
  \bibinfo{pages}{246403} (\bibinfo{year}{2007}).

\bibitem[{\citenamefont{Sanchez et~al.}(1984)\citenamefont{Sanchez, Ducastelle,
  and Gratias}}]{sanchezCE}
\bibinfo{author}{\bibfnamefont{J.~M.} \bibnamefont{Sanchez}},
  \bibinfo{author}{\bibfnamefont{F.}~\bibnamefont{Ducastelle}},
  \bibnamefont{and} \bibinfo{author}{\bibfnamefont{D.}~\bibnamefont{Gratias}},
  \bibinfo{journal}{Physica A} \textbf{\bibinfo{volume}{128A}},
  \bibinfo{pages}{334} (\bibinfo{year}{1984}).

\bibitem[{\citenamefont{Zunger}(1994)}]{Zunger}
\bibinfo{author}{\bibfnamefont{A.}~\bibnamefont{Zunger}}, in
  \emph{\bibinfo{booktitle}{Statics and Dynamics of Alloy Phase
  Transformations, NATO ASI}} (\bibinfo{year}{1994}), p. \bibinfo{pages}{361}.

\bibitem[{\citenamefont{van~de Walle and Ceder}(2002)}]{atat1}
\bibinfo{author}{\bibfnamefont{A.}~\bibnamefont{van~de Walle}}
  \bibnamefont{and} \bibinfo{author}{\bibfnamefont{G.}~\bibnamefont{Ceder}},
  \bibinfo{journal}{J. Phase Equil.} \textbf{\bibinfo{volume}{23}},
  \bibinfo{pages}{348} (\bibinfo{year}{2002}).

\bibitem[{\citenamefont{van~de Walle and Asta}(2002)}]{atat2}
\bibinfo{author}{\bibfnamefont{A.}~\bibnamefont{van~de Walle}}
  \bibnamefont{and} \bibinfo{author}{\bibfnamefont{M.}~\bibnamefont{Asta}},
  \bibinfo{journal}{Modelling Simul. Mater. Sci. Eng.}
  \textbf{\bibinfo{volume}{10}}, \bibinfo{pages}{521} (\bibinfo{year}{2002}).

\bibitem[{\citenamefont{Lang et~al.}(1981)\citenamefont{Lang, Baer, and
  Cox}}]{Lang}
\bibinfo{author}{\bibfnamefont{J.~K.} \bibnamefont{Lang}},
  \bibinfo{author}{\bibfnamefont{Y.}~\bibnamefont{Baer}}, \bibnamefont{and}
  \bibinfo{author}{\bibfnamefont{P.~A.} \bibnamefont{Cox}},
  \bibinfo{journal}{J. Phys. F: Metal Phys.} \textbf{\bibinfo{volume}{11}},
  \bibinfo{pages}{121} (\bibinfo{year}{1981}).

\bibitem[{\citenamefont{Shapira and Reed}(1969)}]{Shapira}
\bibinfo{author}{\bibfnamefont{Y.}~\bibnamefont{Shapira}} \bibnamefont{and}
  \bibinfo{author}{\bibfnamefont{T.~B.} \bibnamefont{Reed}},
  \bibinfo{journal}{J. Appl. Phys.} \textbf{\bibinfo{volume}{40}},
  \bibinfo{pages}{1197} (\bibinfo{year}{1969}).

\bibitem[{\citenamefont{Jayaraman et~al.}(1974)\citenamefont{Jayaraman, Singh,
  Chatterjee, and Devi}}]{Jayaraman}
\bibinfo{author}{\bibfnamefont{A.}~\bibnamefont{Jayaraman}},
  \bibinfo{author}{\bibfnamefont{A.~K.} \bibnamefont{Singh}},
  \bibinfo{author}{\bibfnamefont{A.}~\bibnamefont{Chatterjee}},
  \bibnamefont{and} \bibinfo{author}{\bibfnamefont{S.~U.} \bibnamefont{Devi}},
  \bibinfo{journal}{Phys. Rev. B} \textbf{\bibinfo{volume}{9}},
  \bibinfo{pages}{2513} (\bibinfo{year}{1974}).

\bibitem[{\citenamefont{Goncharenko and Mirebeau}(1998)}]{Goncharenko}
\bibinfo{author}{\bibfnamefont{I.~N.} \bibnamefont{Goncharenko}}
  \bibnamefont{and} \bibinfo{author}{\bibfnamefont{I.}~\bibnamefont{Mirebeau}},
  \bibinfo{journal}{Phys. Rev. Lett.} \textbf{\bibinfo{volume}{80}},
  \bibinfo{pages}{1082} (\bibinfo{year}{1998}).

\bibitem[{not({\natexlab{a}})}]{note:AFMtests}
\bibinfo{note}{For the magnetic ordering tests, most of the structures were
  selected due to their small size and high symmetry, while the largest 8-atom
  compound due to a suspected magnetic instability inferred from the early
  problems in electronic convergence (later linked to an instability of the RMM
  algorithm).}

\bibitem[{not({\natexlab{b}})}]{note-pmfm}
\bibinfo{note}{In the single-sublattice mean-field theory, the PM-FM energy
  difference is equal to $1.5T_c$. According to the argument from the text, the
  $T_c$ for this estimate should be that of pure Eu.}

\bibitem[{\citenamefont{Ducastelle}(1991)}]{DucastelleBook}
\bibinfo{author}{\bibfnamefont{F.}~\bibnamefont{Ducastelle}},
  \emph{\bibinfo{title}{Order and Phase Stability in Alloys}}
  (\bibinfo{publisher}{North Holland}, \bibinfo{address}{Amsterdam},
  \bibinfo{year}{1991}), vol.~\bibinfo{volume}{3}, p. \bibinfo{pages}{142}.

\bibitem[{\citenamefont{H\"aggkvist et~al.}(2007)\citenamefont{H\"aggkvist,
  Rosengren, Lundow, Markstr\"om, Andr\'en, and Kundrotas}}]{Ising}
\bibinfo{author}{\bibfnamefont{R.}~\bibnamefont{H\"aggkvist}},
  \bibinfo{author}{\bibfnamefont{R.}~\bibnamefont{Rosengren}},
  \bibinfo{author}{\bibfnamefont{P.~H.} \bibnamefont{Lundow}},
  \bibinfo{author}{\bibfnamefont{K.}~\bibnamefont{Markstr\"om}},
  \bibinfo{author}{\bibfnamefont{D.}~\bibnamefont{Andr\'en}}, \bibnamefont{and}
  \bibinfo{author}{\bibfnamefont{P.}~\bibnamefont{Kundrotas}},
  \bibinfo{journal}{Adv. Phys.} \textbf{\bibinfo{volume}{56}},
  \bibinfo{pages}{653} (\bibinfo{year}{2007}).

\bibitem[{\citenamefont{Barabash
  et~al.}(2008{\natexlab{a}})\citenamefont{Barabash, Ozolins, and
  Wolverton}}]{LAST}
\bibinfo{author}{\bibfnamefont{S.~V.} \bibnamefont{Barabash}},
  \bibinfo{author}{\bibfnamefont{V.}~\bibnamefont{Ozolins}}, \bibnamefont{and}
  \bibinfo{author}{\bibfnamefont{C.}~\bibnamefont{Wolverton}},
  \bibinfo{journal}{Phys. Rev. Lett.} \textbf{\bibinfo{volume}{101}},
  \bibinfo{pages}{155704} (\bibinfo{year}{2008}{\natexlab{a}}).

\bibitem[{\citenamefont{Barabash and Ozolins}(2010)}]{BAST}
\bibinfo{author}{\bibfnamefont{S.~V.} \bibnamefont{Barabash}} \bibnamefont{and}
  \bibinfo{author}{\bibfnamefont{V.}~\bibnamefont{Ozolins}},
  \bibinfo{journal}{Phys. Rev. B} \textbf{\bibinfo{volume}{81}},
  \bibinfo{pages}{075212} (\bibinfo{year}{2010}).

\bibitem[{\citenamefont{Barabash
  et~al.}(2008{\natexlab{b}})\citenamefont{Barabash, Ozolins, and
  Wolverton}}]{strainLAST}
\bibinfo{author}{\bibfnamefont{S.~V.} \bibnamefont{Barabash}},
  \bibinfo{author}{\bibfnamefont{V.}~\bibnamefont{Ozolins}}, \bibnamefont{and}
  \bibinfo{author}{\bibfnamefont{C.}~\bibnamefont{Wolverton}},
  \bibinfo{journal}{Phys. Rev. B} \textbf{\bibinfo{volume}{78}},
  \bibinfo{pages}{214109} (\bibinfo{year}{2008}{\natexlab{b}}).

\bibitem[{\citenamefont{Catlett and et~al.}(2007)}]{TeraGrid}
\bibinfo{author}{\bibfnamefont{C.}~\bibnamefont{Catlett}} \bibnamefont{and}
  \bibinfo{author}{\bibnamefont{et~al.}}, \emph{\bibinfo{title}{HPC and Grids
  in Action}} (\bibinfo{publisher}{IOS Press}, \bibinfo{address}{Amsterdam},
  \bibinfo{year}{2007}).

\end{thebibliography}

\end{document}